%% file: main.tex
\documentclass[12pt,a4paper]{article}
\usepackage{graphicx,amssymb,amsmath,color,scalefnt}
\usepackage{placeins}
\usepackage{multirow, makecell}
\usepackage{soul}
\usepackage{chngpage}
\usepackage{geometry}
\usepackage[dvipsnames]{xcolor}
\usepackage{rotating}
\usepackage[normalem]{ulem}
\usepackage{jheppub}
\usepackage{scalerel}
\usepackage{makecell}
\usepackage{appendix}
\usepackage{stackengine}
\usepackage{booktabs}
\renewcommand{\thefootnote}{\fnsymbol{footnote}}

\makeatletter\g@addto@macro\bfseries{\boldmath}\makeatother

\DeclareMathAlphabet{\mathsfit}{\encodingdefault}{\sfdefault}{m}{sl}

\newcommand{\FDF}[1][]{\varphi^\dagger #1\!\overleftrightarrow{D}\!_\mu\varphi}
\newcommand{\FDFI}[1][]{\varphi^\dagger #1\!\overleftrightarrow{D}^I\!\!\!_\mu\:\varphi}
\newcommand{\mailto}[1]{\href{mailto:#1}{#1}}

\begin{document}

\title{Future collider constraints on top-quark operators}

\author[a,b]{Fernando Cornet-Gomez\footnote[1]{\mailto{fernando.cornet@uco.es}},}
\author[c]{V\'ictor Miralles\footnote[2]{\mailto{victor.miralles@manchester.ac.uk}},}%
\author[d]{Marcos Miralles L\'opez\footnote[3]{\mailto{marcos.miralles.lopez@cern.ch}},}%
\author[e]{Mar\'ia Moreno Ll\'acer\footnote[4]{\mailto{maria.moreno@ific.uv.es}},}%
\author[e]{Marcel Vos\footnote[5]{\mailto{marcel.vos@cern.ch}}\,}

\affiliation[a]{Departamento de F\'isica, Universidad de C\'ordoba, Campus Universitario de Rabanales, Ctra. N-IV Km. 396, E-14071 C\'ordoba, Spain}
\affiliation[b]{Case Western Reserve U., 10900 Euclid Avenue, Cleveland, OH 44106 USA, United States}
\affiliation[c]{Department of Physics and Astronomy, University of Manchester, Oxford Road, Manchester M13 9PL, United Kingdom}
\affiliation[d]{School of Physics and Astronomy, University of Glasgow, Glasgow G12 8QQ, Scotland, United Kingdom}
\affiliation[e]{IFIC, Universitat de Val\`encia and CSIC, Calle Catedrático José Beltrén 2, E-46980 Paterna, Spain}

\def\TeV{\ifmmode {\mathrm{Te\kern -0.1em V}}\else
                   \textrm{Te\kern -0.1em V}\fi\,}%
\def\GeV{\ifmmode {\mathrm{Ge\kern -0.1em V}}\else
                   \textrm{Ge\kern -0.1em V}\fi\,}%
\def\MeV{\ifmmode {\mathrm{Me\kern -0.1em V}}\else
                   \textrm{Me\kern -0.1em V}\fi\,}%
\def\keV{\ifmmode {\mathrm{ke\kern -0.1em V}}\else
                   \textrm{ke\kern -0.1em V}\fi\,}%
\def\eV{\ifmmode  {\mathrm{e\kern -0.1em V}}\else
                   \textrm{e\kern -0.1em V}\fi\,}%
\let\tev=\TeV
\let\gev=\GeV
\let\mev=\MeV
\let\kev=\keV
\let\ev=\eV

\def\iab{\mbox{ab$^{-1}$}}
\def\ifb{\mbox{fb$^{-1}$}}
\def\ipb{\mbox{pb$^{-1}$}}
\def\inb{\mbox{nb$^{-1}$}}

\definecolor{Purple}{rgb}{0.6, 0.4, 0.8}
\newcommand{\vm}[1]{{ \color{orange} #1}}
\newcommand{\vmcom}[1]{{ \bf \color{Purple} #1}}
\newcommand{\com}[1]{{ \color{red} #1}}
\newcommand{\myComment}[1]{}
\newcommand{\mv}[1]{\textcolor{green!70!black}{#1}}
\newcommand{\fc}[1]{\textcolor{blue}{FCG: #1}}

\newcommand{\HEPfit}{\texttt{HEPfit }}
\newcommand{\ttbar}{{\ensuremath{t\bar{t}}}}

\renewcommand{\thefootnote}{\arabic{footnote}}

\abstract{
In this paper we present updated constraints on the top-quark sector of the Standard Model Effective Field Theory using data available from Tevatron, LEP and the LHC. Bounds are obtained for the Wilson coefficients from a global fit including the relevant two-fermion operators, four-quark operators and two-quark two-lepton operators. We compare the current bounds with the prospects for the high luminosity phase of the Large Hadron Collider and future lepton colliders. 
}

\maketitle

\section{Introduction}

The Standard Model Effective Field Theory (SMEFT)~\cite{Buchmuller:1985jz} provides a framework to order and interpret the wide variety of measurements of Standard Model (SM) processes at collider experiments. Measurements of top-quark production at the Tevatron and the LHC provide important constraints on the Wilson coefficients for operators involving heavy quarks~\cite{Buckley:2015nca,Buckley:2015lku,Brivio:2019ius,Durieux:2019rbz,Bissmann:2019qcd,Hartland:2019bjb,Ellis:2020unq,Ethier:2021bye,Miralles:2021dyw,Alasfar:2022zyr,Aoude:2022aro,Bartocci:2024fmm,Maltoni:2024dpn,Celada:2024mcf,Miralles:2024huv,CMS:2024kvw,terHoeve:2025gey}
. 

Precision measurements of differential cross sections and asymmetries~\cite{Rosello:2015sck} in top-quark pair production provide bounds on the top-gluon vertex and $q\bar{q}t\bar{t}$ operators. Measurements of single top-quark production and top-quark decay constrain the $tWb$ vertex. Run 2 of the LHC opened up rare associated production processes that yield direct bounds on the top-quark couplings to neutral bosons, i.e.~the $t\bar{t} \gamma$, $t\bar{t}Z$ and $t\bar{t}H$ vertices. The observation of four-top-quark production \cite{ATLAS:2021kqb,ATLAS:2023ajo,CMS:2023ftu} and studies of $t\bar{t} b\bar{b}$ \cite{CMS:2020grm} even yield the first direct constraints on four-heavy-quark  operators. The least constrained set of operators is formed by two-lepton-two-heavy-quark ($l^{+}l^{-}Q\bar{Q}$) operators. There are five independent $l^{+}l^{-}Q\bar{Q}$  operators involving top quarks (15 if lepton flavour universality is dropped) that are only very loosely constrained by measurements of $pp \rightarrow t\bar{t}l^+l^-$ production~\cite{CMS:2023xyc,ATLAS:2025yww}.  

Projections of the high-luminosity phase of the LHC (HL-LHC) envisage a substantial improvement of the current bounds~\cite{Durieux:2019rbz}, taking advantage of the large data set to bring the boosted production regime and exploration of rare production processes into the realm of precision physics. Future measurements at high-energy lepton colliders operated above the top-quark pair production threshold can improve the bounds for operators affecting the top-quark electro-weak interactions and in particular the two-lepton-two-heavy-quark operators~\cite{Durieux:2018tev}.

In this paper, we update and complete the studies reported in Refs.~\cite{Durieux:2018tev, Durieux:2019rbz,Durieux:2022cvf,deBlas:2022ofj,Schwienhorst:2022yqu}. We study new LHC measurements of the entanglement of top-quark pairs~\cite{CMS:2024pts,CMS:2024zkc} and include projections for $t\bar{t}l^+l^-$ for the HL-LHC. We adopt up-to-date scenarios for all future electron-positron collider projects, and include a study of the potential of a muon collider. 

In this work, we focus into the high energy collider constraints so we leave the study of constraints from low-energy processes to future works. We would like to emphasise here that these constraints could be relevant for some of the operators \cite{Bissmann:2019gfc,Bissmann:2020mfi,Bruggisser:2021duo,Bruggisser:2022rhb,Grunwald:2023nli,Garosi:2023yxg,Allwicher:2023shc,Bartocci:2023nvp,Cirigliano:2023nol,Gisbert:2024sjw,Mantani:2025bqu} and that developments on this side are currently underway by some of the authors of this work \cite{deBlas:2025xhe}. 

During the last years, there has been an important development of computing tools in the SMEFT community that allows for automatisation of the matching \cite{Criado:2017khh,DasBakshi:2018vni,Fuentes-Martin:2022jrf,Carmona:2021xtq,Chala:2024llp,Guedes:2023azv,Guedes:2024vuf}, the running \cite{Aebischer:2018bkb,Fuentes-Martin:2020zaz,DiNoi:2022ejg}, the prediction of the collider observables \cite{Degrande:2020evl,Brivio:2020onw,Dedes:2023zws,Allwicher:2022mcg}, and the performance of global fits \cite{DeBlas:2019ehy,Aebischer:2018iyb,Straub:2018kue,EOSAuthors:2021xpv,Ellis:2020unq,Giani:2023gfq} -- look at Ref.~\cite{Aebischer:2023nnv} for a recent review. For the case of this paper, all the fits shown have been obtained using the \texttt{HEPfit} code \cite{DeBlas:2019ehy} which employs a Markov Chain Monte Carlo algorithm whose implementation is based on the \texttt{Bayesian Analysis Toolkit}~\cite{Caldwell:2008fw}. The versatility and efficiency of \texttt{HEPfit} allows performing a global fit in the SMEFT \cite{Durieux:2019rbz,Miralles:2021dyw,Miralles:2024huv} on top of the SM~\cite{deBlas:2016ojx, deBlas:2021wap, deBlas:2022hdk} and particular new physics (NP) extensions \cite{Coutinho:2024zyp,Karan:2023kyj,Eberhardt:2021ebh}.

This work is organised as follows. In section~\ref{sec:smeft_basis} we describe the operators included in our fit as well as the theoretical assumptions. In section~\ref{sec:current} we described the inputs and results for the fit including the data currently available. In section~\ref{sec:hl-lhc} we describe the prospects that we have included for the HL-LHC and we show the expected improvement on the LHC by the final stage of the HL-LHC. In section~\ref{sec:ee} we describe the expected observables in the presence of a $e^+e^-$ collider as well as comparing the different scenarios. In section~\ref{sec:muon} we study the projections of a future muon collider. Finally, in section~\ref{sec:conclusion}, we summarise the results, highlighting the most relevant findings from our study.

\section{SMEFT basis}
\label{sec:smeft_basis}

The decoupling theorem \cite{Appelquist:1974tg} guarantees that the contributions of the heavy degrees of freedom to the physical amplitudes are suppressed by the inverse of their masses up to logarithmic corrections.\footnote{Strictly speaking, this does not generally hold in spontaneously broken gauge theories when the masses of the heavy particles arise from the symmetry breaking itself. In such cases, non-decoupling effects may occur.} This allows us to condensate the NP effects into higher dimensional operators built with the SM particle content. In the SMEFT the NP is assumed to be invariant under the same gauge symmetries as the SM Lagrangian before the electroweak symmetry breaking. Imposing also that the accidental symmetries of the SM are conserved (baryon and lepton number) the SMEFT Lagrangian can be written as
\begin{equation}
\mathcal{L}_\text{eff} = \mathcal{L}_\text{SM} +   \left( \frac{1}{\Lambda^2} \sum_i C_i O_i + \text{h.c.} \right)  + \mathcal{O}\left(\Lambda^{-4} \right) ,
\end{equation}
where the operators $O_i$ are built from SM fields and the coefficients $C_i$ are known as the Wilson coefficients (WC). 

In this work, we will assume charge-parity conservation in the NP and study a subset of WC that are relevant in top-quark physics. These operators are shown in Tab.~\ref{tab:WilsonDefined} and defined in Tab.~\ref{tab:WilsonOperators}. We will use a linear combination of the WC of the Warsaw basis \cite{Grzadkowski:2010es} following the prescription of the LHC top-quark WG \cite{AguilarSaavedra:2018nen}. In our Lagrangian we assume a flavour symmetry $U(2)^5$ so we only distinguish the third generation \cite{Faroughy:2020ina}. 

\begin{table}[htp!]
\centering
\renewcommand{\arraystretch}{1.2}
\begin{tabular*}{\textwidth}{@{\extracolsep{\fill}}|@{\,}c@{\,}|@{}c@{\,}|@{}c@{\,}|@{}c@{\,}| } 
 \hline
 \multicolumn{4}{|c|}{\textbf{Coefficients fitted}} \\
 \hline
  \multirow{3}{*}{2-quark} & $C_{ t G}$  & $C_{\varphi Q}^3$  & $C_{\varphi Q}^- = C_{\varphi Q}^1-C_{\varphi Q}^3$  \\ 
    &   $C_{\varphi t}$ &  $C_{\varphi b}$  &    $C_{t Z} = c_W C_{t W}-s_W C_{t B}$   \\ 
 
    &  -- &  $C_{t \varphi }$ &   $C_{t W}$    \\ 
 \hline
 \hline
     \multirow{5}{*}{4-quark} &  $ C_{tu}^8 = \sum\limits_{\scaleto{i=1,2}{4pt}} 2C_{uu}^{(i33i)}$ &  $ C_{td}^8 = \sum\limits_{\scaleto{i=1,2}{4pt}} C_{ud}^{8(33ii)}$ &  $C_{Qq}^{1,8} = \sum\limits_{\scaleto{i=1,2}{4pt}} C_{qq}^{1(i33i)}+3C_{qq}^{3(i33i)}$  \\

        & $ C_{Qu}^{8} = \sum\limits_{\scaleto{i=1,2}{4pt}} C_{qu}^{8(33ii)}$ &  $ C_{Qd}^{8} =\sum\limits_{\scaleto{i=1,2}{4pt}} C_{qd}^{8(33ii)}$  &   $C_{Qq}^{3,8} =\sum\limits_{\scaleto{i=1,2}{4pt}} C_{qq}^{1(i33i)}-C_{qq}^{3(i33i)}$  \\

    &    $C_{tq}^{8} = \sum\limits_{\scaleto{i=1,2}{4pt}} C_{uq}^{8(ii33)}$ &  $ C_{td}^1 = \sum\limits_{\scaleto{i=1,2}{4pt}} C_{ud}^{1(33ii)}$  &   $ C_{tu}^1 = \sum\limits_{\scaleto{i=1,2}{4pt}} C_{uu}^{(ii33)} + \frac{1}{3}C_{uu}^{(i33i)}$ \\
     & $ C_{Qu}^{1} = \sum\limits_{\scaleto{i=1,2}{4pt}} C_{qu}^{1(33ii)}$ &  $ C_{Qd}^{1} =\sum\limits_{\scaleto{i=1,2}{4pt}} C_{qd}^{1(33ii)}$  &   $C_{Qq}^{1,1} =\sum\limits_{\scaleto{i=1,2}{4pt}} C_{qq}^{1(ii33)}+\frac{1}{6}C_{qq}^{1(i33i)}+\frac{1}{2}C_{qq}^{3(i33i)}$  \\

    &    -- &  $C_{tq}^{1} = \sum\limits_{\scaleto{i=1,2}{4pt}} C_{uq}^{1(ii33)}$  &   $ C_{Qq}^{3,1} = \sum\limits_{\scaleto{i=1,2}{4pt}} C_{qq}^{3(ii33)} + \frac{1}{6}\big(C_{qq}^{1(i33i)}-C_{qq}^{3(i33i)}\big)$ \\

 \hline
 \hline
 \multirow{3}{*}{\makecell{2-quark\\2-lepton}}  & $C_{eb}$ & $C_{et}$ & $ C_{ l Q}^+ = C_{lQ}^1+C_{lQ}^3$  \\
 
   & $C_{lb}$ &  $C_{lt}$ &  $C_{ l Q}^- = C_{lQ}^1-C_{lQ}^3$  \\
  
    &  -- & --  &   $C_{eQ}$ \\
   \hline
\end{tabular*}
\caption{Here we present the Wilson coefficients that have been fitted in our analysis in terms of those of Tab.~\ref{tab:WilsonOperators}. Those in  first block are related with the two-quark operators, those in the second block with the four-quark operators and the last block is related with the two-quark two-lepton operators.}
\label{tab:WilsonDefined}
\end{table}

\begin{table}[!htb]
\centering
\renewcommand{\arraystretch}{1.2}
\begin{tabular*}{\textwidth}{@{\extracolsep{\fill}} |c|c@{\quad}||c@{\quad}|c@{\quad}| } 
 \hline
 \multicolumn{4}{|c|}{\textbf{Relevant operators}} \\
 \hline
 Coefficient & Operator & Coefficient & Operator \\ 
 \hline
$C_{\varphi Q}^1 $
		&   
		$\left(\bar{ Q} \gamma^\mu  Q \right) \left(\FDF[i] \right)  $ &$ C_{\varphi Q}^3$
		&$ 
		 \left(\bar{ Q} \tau^I \gamma^\mu  Q \right) \left(\FDFI[i] \right)$ \\ 
 $C_{\varphi t}$
		&
		$\left(\bar{ t}\gamma^\mu  t\right)\left(\FDF[i]\right) $& $C_{\varphi b}$
		& $
		\left( \bar{ b}\gamma^\mu  b \right)\left( \FDF[i]\right)$ \\ 
	     $C_{t\varphi}$
		&  
		$\left( \bar{ Q}  t \right)
		\left( \epsilon\varphi^* \; \varphi^\dagger\varphi \right) $ 
        &
        $C_{tG}$ & $\left(\bar{t} \sigma^{\mu \nu} T^{A} t \right)\left(\epsilon \varphi^* G_{\mu \nu}^{A} \right)$\\
	
	$C_{tW}$
		& 
		$\left(\bar{ Q}\tau^I\sigma^{\mu\nu}  t  \right)\left(\epsilon\varphi^* W_{\mu\nu}^I  \right) $&
		$ C_{tB} $ 
		&
		$	\left(\bar{ Q}\sigma^{\mu\nu}  t \right)
	 \left( \epsilon\varphi^* B_{\mu\nu} \right)$\\
 \hline
 \hline
 $C_{qq}^{1(ijkl)}$
		&  
		$(\bar{ q}_i\gamma^\mu  q_j) (\bar{  q}_k\gamma_\mu  q_l)$ &
 $C_{qq}^{3(ijkl)}$
		&  
		$(\bar{ q}_i\tau^I\gamma^\mu  q_j) (\bar{  q}_k\tau^I\gamma_\mu  q_l)$ \\

 $C_{ud}^{8(ijkl)}$
		&  
		$(\bar{ u}_i\gamma^\mu T^A u_j) (\bar{  d}_k\gamma_\mu T^A d_l)$ &
         $C_{qu}^{8(ijkl)}$
		&  
		$(\bar{ q}_i\gamma^\mu T^A q_j) (\bar{  u}_k\gamma_\mu T^A u_l)$  \\

		 $C_{qd}^{8(ijkl)}$ &  
		$(\bar{ q}_i\gamma^\mu T^A q_j) (\bar{  d}_k\gamma_\mu T^A d_l)$  &  $C_{uu}^{(ijkl)}$
		&  
		$(\bar{ u}_i\gamma^\mu  u_j) (\bar{  u}_k\gamma_\mu  u_l)$\\

 $C_{ud}^{1(ijkl)}$
		&  
		$(\bar{ u}_i\gamma^\mu  u_j) (\bar{  d}_k\gamma_\mu  d_l)$ &
         $C_{qu}^{1(ijkl)}$
		&  
		$(\bar{ q}_i\gamma^\mu  q_j) (\bar{  u}_k\gamma_\mu  u_l)$  \\

 $C_{qd}^{1(ijkl)}$
		&  
		$(\bar{ q}_i\gamma^\mu  q_j) (\bar{  d}_k\gamma_\mu  d_l)$ &-- & -- \\
 \hline
 \hline

	$C_{lQ}^1$
		&$  \left(\bar{ Q}\gamma_\mu  Q\right)	\left( \bar{ l}\gamma^\mu  l \right)$ &
	$C_{lQ}^3$
		& $ \left( \bar{ Q}\tau^I\gamma_\mu  Q \right)	\left(\bar{ l}\tau^I\gamma^\mu  l \right) $
	\\
	$C_{lt} $&$ \left( \bar{ t}\gamma_\mu  t\right) 	 \left(\bar{ l}\gamma^\mu  l \right)$ &
	$C_{lb} $&$  \left( \bar{ b}\gamma_\mu  b\right)		\left( \bar{ l}\gamma^\mu  l\right) $
	\\
	$C_{eQ} $&$ \left( \bar{ Q}\gamma_\mu  Q \right) 	\left(\bar{ e}\gamma^\mu  e \right)$&
	$C_{et} $& $ \left(\bar{ t}\gamma_\mu  t \right) 		\left( \bar{ e}\gamma^\mu  e \right)$
	\\
	$C_{eb} $& $ \left(\bar{ b}\gamma_\mu  b \right) 		\left( \bar{ e}\gamma^\mu  e \right)$& -- & -- \\
	\hline
\end{tabular*}
\caption{Here we show the most relevant operators whose linear combinations have been fitted in this work. The first block are two-quark operators, the second block are four-quark operators and the last block are two-quark two-lepton operators. In these operators $Q$ is the left-handed doublet of the two heaviest quarks, the Latin letters are flavour indices, $\tau^I$ are the Pauli matrices, $T^A=\lambda^A/2$ with $\lambda^A$ the Gell-Mann matrices. In the lepton fields we consider both light leptons, electrons and muons, with the same WC for both of them.
}
\label{tab:WilsonOperators}
\end{table}

The SMEFT contribution to a physical observable can therefore be written as
\begin{equation}
    X_{\text{SMEFT}}=X_{\text{SM}}+\sum_i \frac{C_i}{\Lambda^2}X^{\rm{int}}_i+\sum_{ij}\frac{C_i C_j}{\Lambda^4} X^{\rm{quad}}_{ij}+\mathcal{O}(\Lambda^{-4}) \quad.
\end{equation}
The linear terms arise from the interference of the NP contributions with the SM and the quadratic terms come from squaring the NP contributions. In our work we will only include the leading-order (LO) contribution from the dimension-six operators. The quadratic terms are of order $\Lambda^{-4}$, the same order as the interference of the dimension-eight operators with the SM. A complete treatment of the $\mathcal{O}(\Lambda^{-4})$ terms is currently not possible. Several groups have developed recipes to truncate the expansion and define uncertainties to ensure validity of the SMEFT~\cite{Brivio:2022pyi}, but there is no consensus on which approach to adopt. In our analysis, we include only the linear terms proportional to $\Lambda^{-2}$. In analyses of the LHC and HL-LHC projections, quadratic terms can be sizeable, and our results are therefore conservative in comparison with other groups. Quadratic terms generally play a minor role at lepton colliders at energies up to several \tev{}. Therefore, our projections for the Higgs factory would not be altered significantly by the inclusion of $\Lambda^{-4}$ terms. 

The observables included in the fit are measurements at the LHC, Tevatron and LEP. In order to parametrise the dependence of the observable on the Wilson coefficients we employ the Monte Carlo generator \texttt{MadGraph5}\_\texttt{aMC@NLO} \cite{Alwall:2014hca} using two Universal Feynrules Models \cite{Darme:2023jdn}: \texttt{SMEFTsim} \cite{Brivio:2020onw} and \texttt{SMEFT@NLO} \cite{Degrande:2020evl}. From these models we obtain the linear contribution of the different WC to the LHC processes included in the fit. For all the current LHC data we include the next-to-LO QCD corrections in the matrix elements, using  \texttt{SMEFT@NLO}. Only for some of the additional observables expected at the HL-LHC, like the $pp \rightarrow t\bar{t}l^+l^-$ process, we use \texttt{SMEFTsim}.

\section{Current Constraints}
\label{sec:current}

\begin{table*}[h!]
\centering
\resizebox{\textwidth}{!}{%
\begin{tabular}{|l|c|c|c|c|c|c|}
\hline
Process & Observable & $\sqrt{s}$  & $L_\text{int}$  & Experiment & SM  & Ref.\\ \hline
$pp \rightarrow \ttbar $ & $d\sigma/dm_\ttbar$ (15 bins) & 13 TeV & 137~\ifb & CMS & \cite{Czakon:2013goa} & \cite{CMS:2021vhb} \\
$pp \rightarrow \ttbar $ & $dA_C/dm_\ttbar$ (5 bins) & 13 TeV & 139~\ifb & ATLAS & \cite{Czakon:2013goa} & \cite{ATLAS:2022waa} \\
$pp \rightarrow \ttbar $ & $D (m_\ttbar \sim 2 m_t)$  & 13 TeV & 137~\ifb & CMS & MG5 & \cite{CMS:2024pts}  \\
$pp \rightarrow \ttbar $ & $D_n (m_\ttbar > 0.8~\tev)$  & 13 TeV & 137~\ifb & CMS & MG5 & \cite{CMS:2024zkc}  \\
$p p \rightarrow t \bar{t} H$ & $d\sigma/dp_T^H$ (6 bins) & 13 TeV & 139~\ifb & ATLAS &
\cite{deFlorian:2016spz} &  \cite{ATLAS:2022vkf}\\ 
$p p \rightarrow t \bar{t} Z$ &  $d\sigma/dp_T^Z$ (8 bins) & 13 TeV &  140~\ifb  & ATLAS &
\cite{Broggio:2019ewu} & \cite{ATLAS:2023eld} \\ 
$p p \rightarrow t \bar{t} \gamma$ & $d\sigma/dp_T^\gamma$ (10 bins) & 13 TeV & 140~\ifb & ATLAS &
\cite{Bevilacqua:2018woc,Bevilacqua:2018dny} & \cite{ATLAS:2024hmk} \\ 
 $p p \rightarrow t \bar{t} W$ & $\sigma$ & 13 TeV & 138~\ifb  & CMS &
\cite{Buonocore:2023ljm} &  \cite{CMS:2022tkv}  \\
$p p \rightarrow tZq$ & $\sigma$ & 13 TeV & 138~\ifb{}  & CMS &
 \cite{CMS:2021ugv} & \cite{CMS:2021ugv} \\ 
$p p \rightarrow t\gamma q$ & $\sigma$ & 13 TeV & 140~\ifb{}  & ATLAS &
 \cite{ATLAS:2023qdu} & \cite{ATLAS:2023qdu} \\
 $p p \rightarrow t\bar{b}$ (s-ch) & $\sigma$ & 8~\tev{} & 20~\ifb{} & LHC &
\cite{Aliev:2010zk,Kant:2014oha} & \cite{Aaboud:2019pkc} \\ 
$p p \rightarrow tW$ & $\sigma$ & 8~\TeV & 20~\ifb{}  & LHC &
 \cite{Kidonakis:2010ux} &  \cite{Aaboud:2019pkc}
 \\ 
$p p \rightarrow tq$ (t-ch) & $\sigma$ & 8/13~\tev{} & 20/140~\ifb{} & LHC/ATLAS &
\cite{Aliev:2010zk,Kant:2014oha} & \cite{Aaboud:2019pkc}/\cite{ATLAS:2024ojr} 
\\ 
$t \rightarrow Wb $ & $F_0$, $F_L$  & 8/13~\tev{} & 20/139~\ifb{} & LHC/ATLAS &
\cite{PhysRevD.81.111503}  & \cite{Aad:2020jvx}/\cite{ATLAS:2022rms} \\
$p\bar{p} \rightarrow t\bar{t}$ & $dA_{FB}/dm_\ttbar$ (4 bins) & 1.96 TeV & 9.7~\ifb & Tevatron & \cite{Czakon:2014xsa, Czakon:2016ckf} & \cite{CDF:2017cvy}\\
$p\bar{p} \rightarrow t\bar{b}$ (s-ch) & $\sigma$ & 1.96~\tev & 9.7~\ifb & Tevatron & \cite{PhysRevD.81.054028} & \cite{CDF:2014uma} \\
$e^{-} e^{+} \rightarrow b \bar{b} $ & $R_{b}$,  $A_{FBLR}^{bb}$ & $\sim$ 91~\GeV & 202.1~\ipb  & 
LEP/SLD &
 - & \cite{ALEPH:2005ab}  \\ \hline
\end{tabular}%
}
\caption{Measurements included in the LHC fit. For each measurement, the process, the observable, the centre-of-mass energy, the integrated luminosity and the experiment/collider are given. The last two columns list the references for the predictions and measurements that are included in the fit. LHC refers to the combination of ATLAS and CMS measurements. In a similar way, Tevatron refers to the combination of CDF and D0 results, and LEP/SLD to different experiments from those two accelerators. }
\label{tab:measurements}
\end{table*}

The SMEFT fit of the top-quark sector is dominated by LHC measurements of top-quark processes in Run~2~\cite{Durieux:2019rbz}. The observables included in our fit are shown in Tab.~\ref{tab:measurements} which include the most relevant processes needed to constrain the top-quark sector. Given that the correlations among the different measurements are not provided by the experimental collaborations, and specially motivated by the fact that in the HL-LHC the statistical uncertainties will be subleading, we only add the measurement of one collaboration for each process, unless a proper combination has been performed. 

An important workhorse is the top-quark pair production process, which has been characterised to excellent precision. The inclusive cross section measurements reach a precision of better than 2\%, about a factor two better than state-of-the-art SM predictions~\cite{Czakon:2013goa}. Importantly, multi-dimensional differential measurements that cover kinematic regimes from the pair production threshold to the boosted regime are available. In this study, the differential cross section and charge asymmetry are included, both as a function of $m_{t\bar{t}}$. These are key to constrain the large number of degrees of freedom that affect top-quark pair production: the operator coefficient $C_{tG}$, that modifies the top-quark's chromo-magnetic dipole moment, and the fourteen operator coefficients $q\bar{q}Q\bar{Q}$ that arise from integrating out massive new mediators with different Lorentz structures. Two recent measurements sensitive to quantum entanglement ($D$ at threshold~\cite{CMS:2024pts} and $D_n$ in the boosted regime~\cite{CMS:2024zkc}) have been included and their impact is discussed in the next section.

A second important data set involves the charged-current interaction. The $tWb$ vertex is constrained by top-quark decay measurements, such as the $W$-boson helicity fractions and measurements in electro-weak single top-quark production. In this area, legacy data from LHC Run 1, the Tevatron experiments and LEP data continue to play a role in the fit. LHC Run 1 data are used for the $W$-helicities~\cite{Aad:2020jvx}. Combined Tevatron results are used for the forward-backward asymmetry~\cite{CDF:2017cvy} and the s-channel~\cite{CDF:2014uma}, that remain competitive. LEP/SLC data on the $Z\rightarrow b\bar{b}$ vertex provide strong constraints on $C^{(1)}_{\varphi q}$ and $C^{(3)}_{\varphi q}$, that affect the top- and bottom-quark couplings to the $Z$-boson.

Two important new results, compared to Ref.~\cite{Durieux:2019rbz}, are the ATLAS $t\bar{t}Z$~\cite{ATLAS:2023eld} and $t\bar{t}\gamma$~\cite{ATLAS:2024hmk} differential cross section measurements. In the former, the precision of the measurement is substantially improved compared to earlier measurements. The latter presents results for the $t\bar{t}\gamma$ production process, where the contribution of photons produced from top-quark decay products is subtracted, providing for a direct comparison to the $ 2 \to 3$ predictions.

\begin{figure}[h!]
\includegraphics[width=1.0\columnwidth]{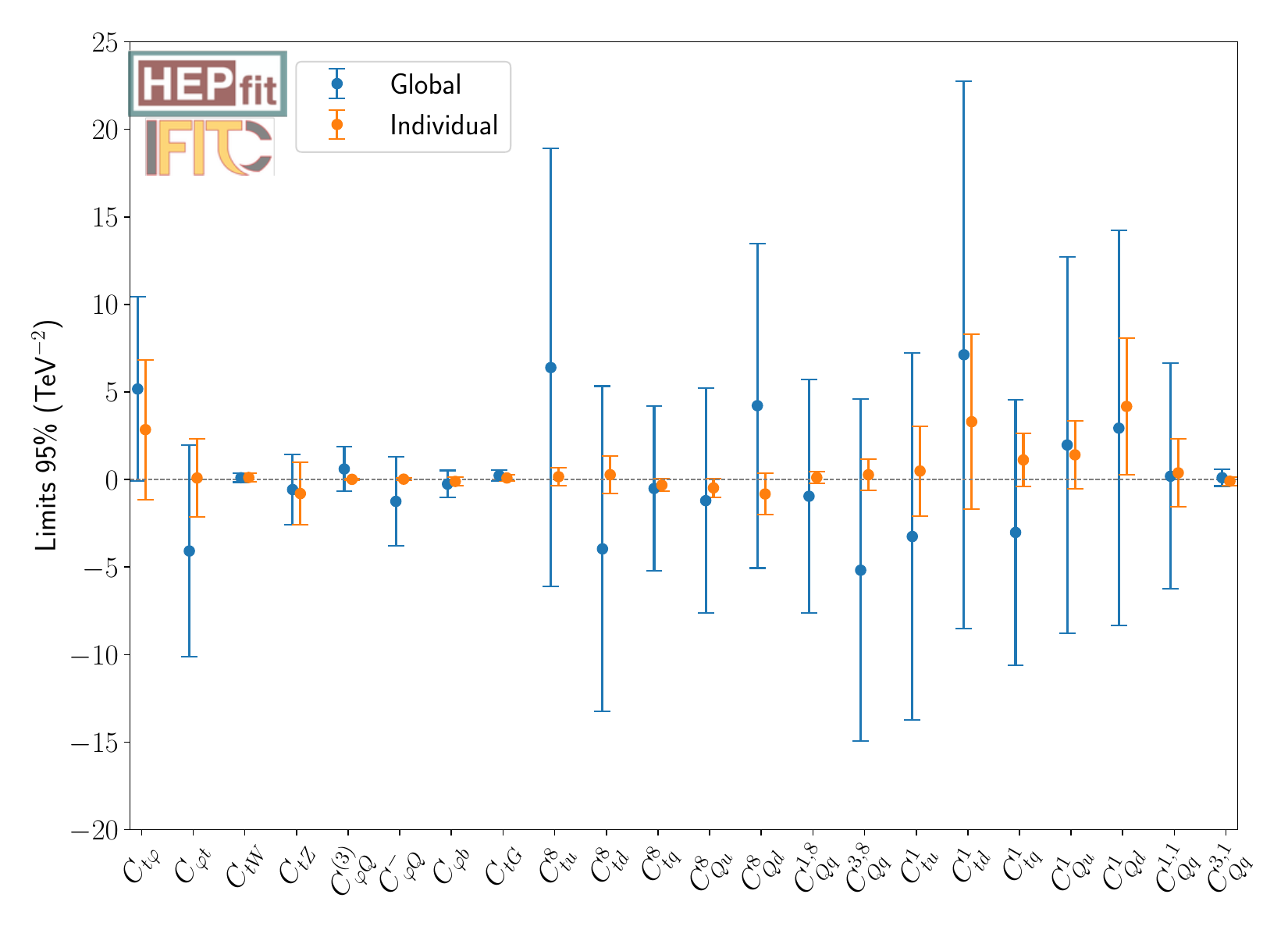}
\caption{\label{fig:lhc_centered_exp}%
The 95\% probability constraints on the WC divided by $\Lambda^2$ using the current LHC data in combination of the legacy data from Tevatron and LEP.
}
\end{figure}

In Fig.~\ref{fig:lhc_centered_exp} and Tab.~\ref{tab:lhc_centered_exp} we show the 95\% probability limits for the top-quark operators using the current LHC data. All Wilson coefficients are found to be compatible with the SM value of 0 at 95\% probability, reflecting the good agreement of the measurements of Tab.~\ref{tab:measurements} with the SM predictions. Regarding the goodness-of-fit, we find that the SM has a $\chi^2$ per degree of freedom of 0.76, whereas the SMEFT fit yields a significantly improved value of 0.56.\footnote{When determining the number of degrees of freedom we have considered each bin as an observable even though this value may be reduced due to correlations among the bins.}

The tightest bounds are obtained on the coefficients of two-fermion operators, where 95\% probability limits range from 0.25 \tev$^{-2}$ for $C_{tW}/\Lambda^{2}$ to 5 \tev$^{-2}$ for $C_{t \varphi}$. These bounds are driven by precise differential cross section measurements in top-quark decay, single top-quark production and associated production processes $t\bar{t}H$, $t\bar{t}Z$ and $t\bar{t}\gamma$. In particular, the studies of rare associated top-quark production processes have evolved from observation to precision measurements. It is illustrative to contrast our result with the first fits of these coefficients at the start of Run 2 in Ref.~\cite{Buckley:2015lku}, where the data still provided insufficient constraints for a global analysis.

\begin{table}[h]
    \centering
    \begin{tabular}{ccc}
        \toprule
        Parameter & Global & Individual \\
        \midrule
        $C_{t\varphi}$  & [-0.1, 10.4] & [-1.1, 6.8] \\
        $C_{\varphi t}$  & [-10.1, 2.0] & [-2.1, 2.3] \\
        $C_{tW}$    & [-0.15, 0.36] & [-0.13, 0.36] \\
        $C_{tZ}$    & [-2.6, 1.4] & [-2.6, 1.0] \\
        $C_{\varphi Q}^3$ & [-0.7, 1.9] & [-0.02, 0.04] \\
        $C_{\varphi Q}^-$ & [-3.8, 1.3] & [-0.04, 0.08] \\
        $C_{\varphi b}$  & [-1.0, 0.5] & [-0.37, 0.15] \\
        $C_{tG}$    & [-0.08, 0.55] & [-0.10, 0.28] \\
        $C_{tu}^8$   & [-6.1, 18.9] & [-0.35, 0.66] \\
        $C_{td}^8$   & [-13.3, 5.3] & [-0.8, 1.3] \\
        $C_{tq}^8$   & [-5.2, 4.2] & [-0.68, 0.04] \\
        $C_{Qu}^8$   & [-7.6, 5.2] & [-1.0, 0.05] \\
        $C_{Qd}^8$   & [-5.1, 13.5] & [-2.0, 0.4] \\
        $C_{Qq}^{1,8}$  & [-7.6, 5.7] & [-0.23, 0.46] \\
        $C_{Qq}^{3,8}$  & [-14.9, 4.6] & [-0.6, 1.1] \\
        $C_{tu}^1$   & [-13.7, 7.2] & [-2.1, 3.0] \\
        $C_{td}^1$   & [-8.5, 22.8] & [-1.7, 8.3] \\
        $C_{tq}^1$   & [-10.6, 4.6] & [-0.4, 2.6] \\
        $C_{Qu}^1$   & [-8.8, 12.7] & [-0.5, 3.4] \\
        $C_{Qd}^1$   & [-8.4, 14.2] & [0.3, 8.1] \\
        $C_{Qq}^{1,1}$  & [-6.3, 6.6] & [-1.5, 2.3] \\
        $C_{Qq}^{3,1}$  & [-0.38, 0.58] & [-0.34, 0.15] \\
        \bottomrule
    \end{tabular}
    \caption{Global marginalised and individual 95\% probability limits using the current LHC data in combination of the legacy data from Tevatron and LEP.}
    \label{tab:lhc_centered_exp}
\end{table}

The bounds on the four-fermion operator coefficients are considerably weaker, with the exception of $C_{Qq}^{3,1}$, which is well constrained by the t-channel single top-quark production \cite{ATLAS:2024ojr}. For the $q\bar{q}t\bar{t}$ operators, individual bounds are usually of the order of one to a few \tev$^{-2}$ and global bounds are typically an order of magnitude worse. The pronounced difference between the marginalised constraints from the global fit and the individual limits is due to unresolved correlations between the coefficients. The Wilson coefficients of $q\bar{q}t\bar{t}$ operators are mainly constrained from measurements in top-quark pair production. There are many degrees of freedom affecting this process, making it difficult to constrain all the coefficients in a global analysis. 
The ``blind directions'' are readily spotted in the correlation matrix for this fit in Fig.~\ref{fig:correlations_LHC_fit}. Tighter bounds are obtained~\cite{Celada:2024mcf} from fits based on parametrisations including terms proportional to $\Lambda^{-4}$.

The fit does not yield meaningful bounds on two-quark two-lepton operators, as the measurements of Tab.~\ref{tab:measurements} have insufficient sensitivity for the four degrees of freedom, and therefore these WC have not been considered for this fit. Indeed, even when setting the NP scale as low as 1 TeV the constraints of these operators in the global fit are too weak to lay within the perturbative regime. Dedicated analyses of $t\bar{t} l^+l^-$ at the LHC with $m_{ll}$ above the $Z$-boson mass window may provide additional constraints on these operators. The analysis of Ref.~\cite{CMS:2023xyc} finds bounds of ${\mathcal{O}}(1~\tev^{-2})$. These bounds, however, strongly rely on terms proportional to $\Lambda^{-4}$. The Run 2 data set cannot constrain these operator coefficients in a global analysis based on a linear parametrisation.

\begin{figure}\centering
\includegraphics[width=1.1\columnwidth]{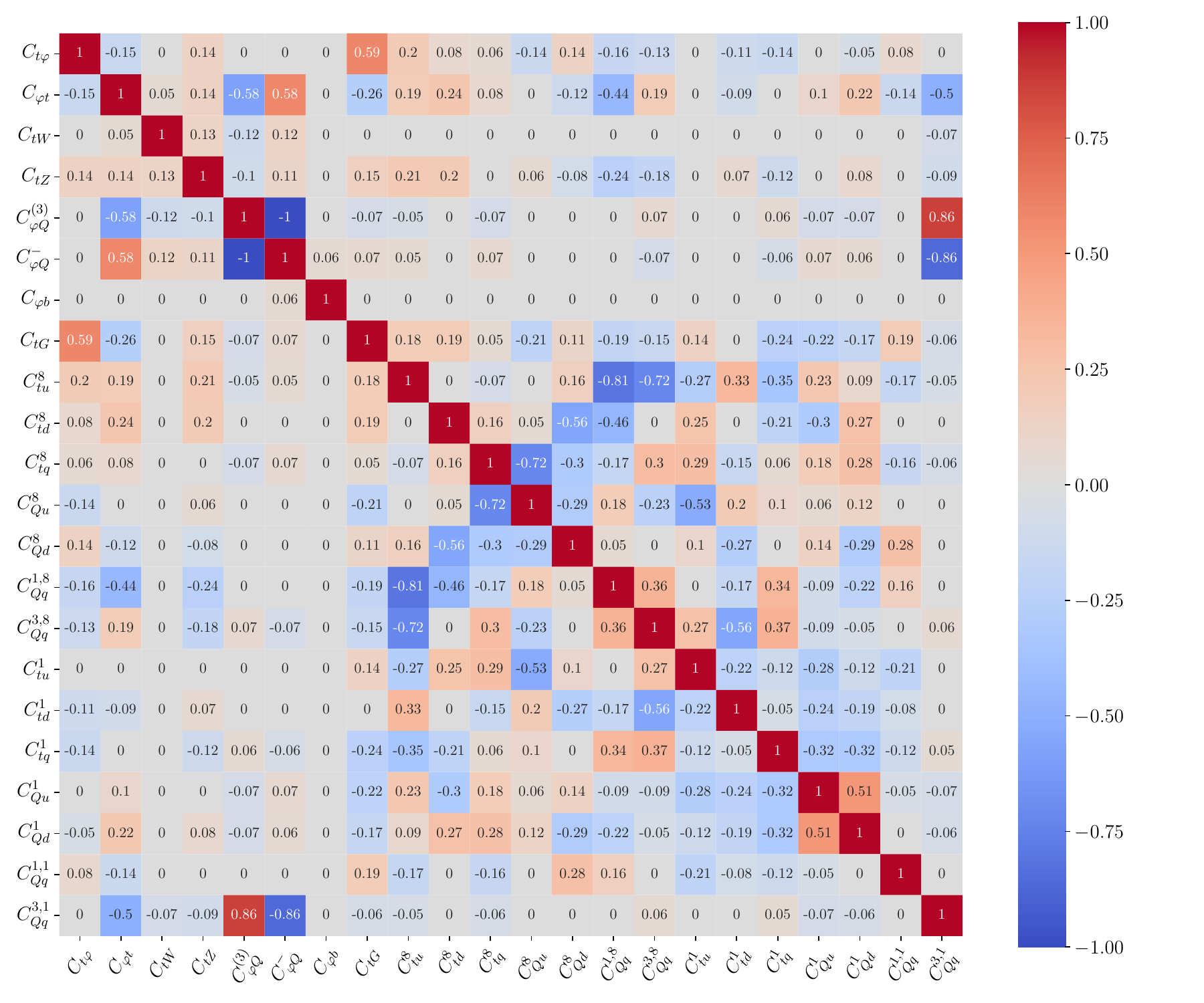}
\caption{\label{fig:correlations_LHC_fit}Correlation matrix obtained for the global fit including the data of the LHC, Tevatron and LEP. Entries smaller than 5\% are set to zero.  }
\end{figure}

Our results are qualitatively in agreement with the results of Ref.~\cite{Celada:2024mcf}. The most significant difference is that our fit sets a much more stringent bound on $C_{tZ}$. This difference can be traced to the differential cross section on $t\bar{t}\gamma$ production \cite{ATLAS:2023qdu}, which has an important constraining power~\cite{Durieux:2019rbz}. Ref.~\cite{Celada:2024mcf} presents a more stringent individual bound on $C_{t \varphi}$, this is likely due to the inclusion of Higgs data beyond $t\bar{t}H$ production.

\subsection{Entanglement measurements}

Top-quark pairs produced at high-energy colliders form a two-qubit system. They are expected to be produced in a quantum-entangled state where the top quark cannot be described without reference to the anti-top quark.  
In Ref.~\cite{Afik:2020onf}, Afik and Muñoz de Nova propose a simple measurement to study top-quark pairs that are formed in a quantum-entangled state at the LHC. Recently, the ATLAS and CMS collaborations indeed observed quantum entanglement in top-quark pair production at the threshold~\cite{ATLAS:2023fsd,CMS:2024pts}. Moreover, the CMS experiment extended the observation to the boosted regime~\cite{CMS:2024zkc}. 
These new measurements have a pronounced sensitivity to $C_{tG}$ and the coefficients of the $q\bar{q}t\bar{t}$ operators in the SMEFT~\cite{Severi:2022qjy}. Adding these measurements can provide a new angle to constrain the top-quark sector. 

The bounds derived from these new measurements are summarised in Fig.~\ref{fig:entanglement_ranking}. For comparison, the ranking also includes the differential charge asymmetry and cross section measurements. For each measurement, two bars are drawn: one that corresponds to the bounds obtained with a linear parametrisation of the $\Lambda^{-2}$ terms (labelled as ``lin.''), and another with a quadratic parametrisation that adds the dimension-six-squared terms of order $\Lambda^{-4}$ (labelled as ``quad.'').  

\begin{figure}[h!]
    \centering
    \includegraphics[width=1.0\linewidth]{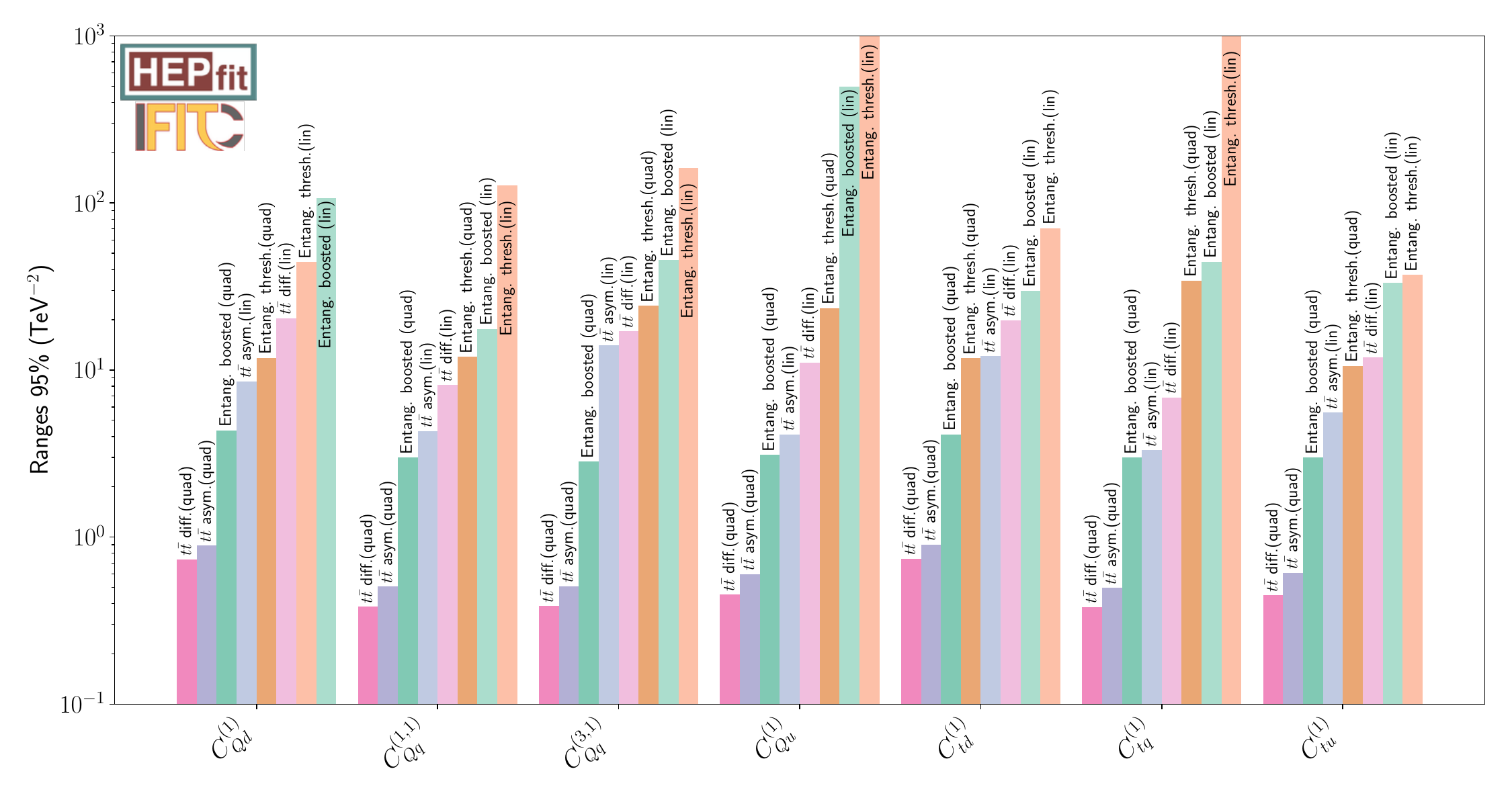}
    \includegraphics[width=1.0\linewidth]{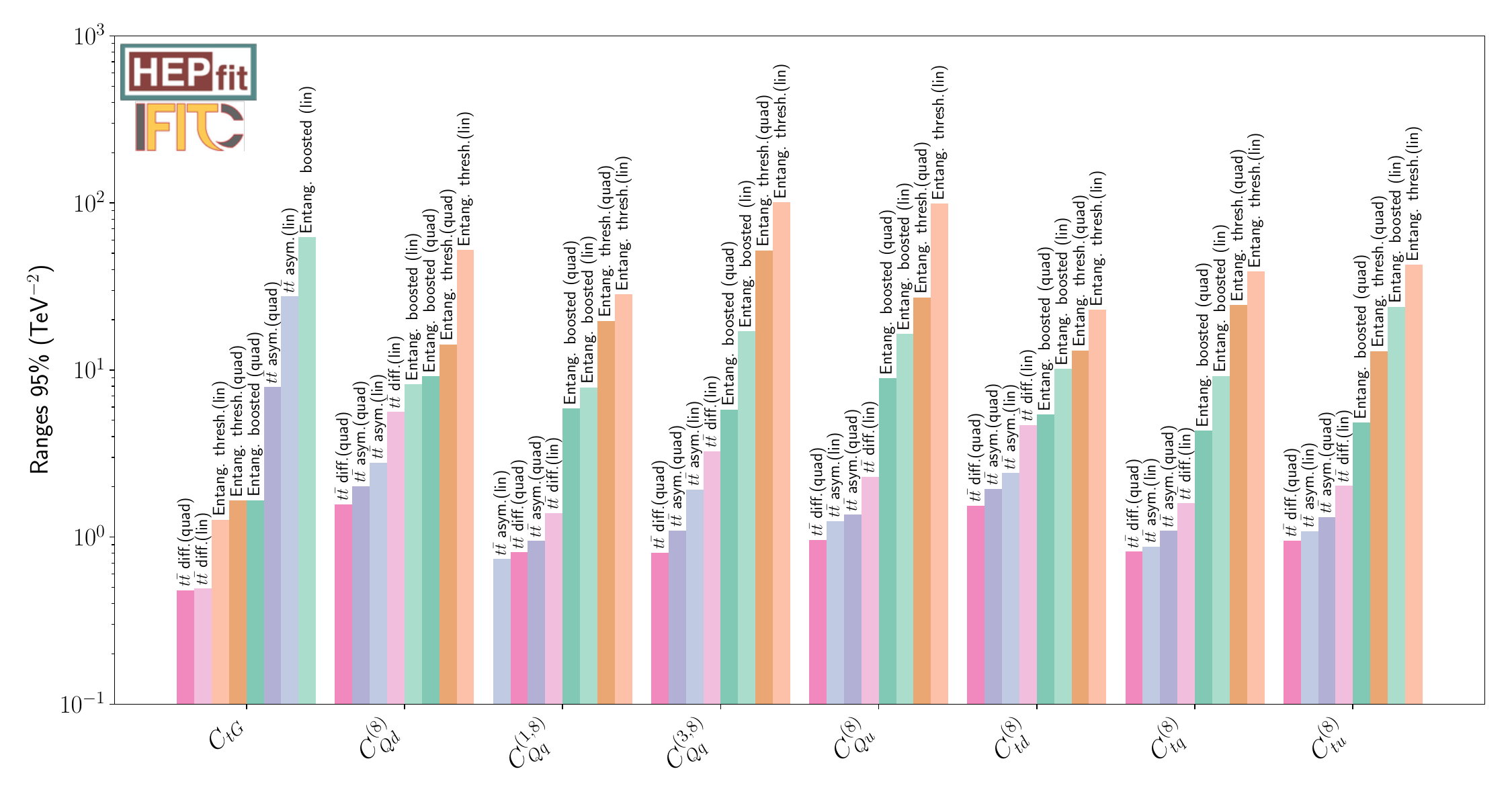}
    \caption{Ranking of the individual 95\% probability bounds on the WC divided by $\Lambda^2$ of $O_{tG}$ and the $q\bar{q}t\bar{t}$ operators. Each measurement in top-quark pair production is shown. }
    \label{fig:entanglement_ranking}
\end{figure}

The measurements of $D$ at the top-quark pair production threshold offer individual bounds on $C_{tG}$ that are of the same order as that of differential cross section measurements. Linear and quadratic bounds are similar in this case. These measurements are currently limited by modelling uncertainties and an improved theory description of the threshold region, including pseudo-bound-state effects, can improve their impact considerably. 

The measurement of $D_n$ is taken as a proxy for the full set of measurements in the boosted regime. The individual bounds derived from the coefficients of $q\bar{q}t\bar{t}$ operators from this measurement in a linear fit are similar to those from associated production processes (such as $t\bar{t}Z$ or $t\bar{t}H$), but cannot compete with the differential cross section and charge asymmetry measurements in top-quark pair production. Bounds including quadratic terms are an order of magnitude better, as terms proportional to $\Lambda^{-4}$ dominate the sensitivity, but the entanglement measurement still ranks behind the charge asymmetry and cross section measurements, which also gain considerable sensitivity from the quadratic terms. 

In Fig.~\ref{fig:entanglement}, the 95\% probability bounds on $C_{tG}$ and the 4-quark operators of the fit to LHC data are shown before and after adding the CMS measurements of $D$ at threshold~\cite{CMS:2024pts} and $D_n$ in the boosted regime~\cite{CMS:2024zkc} to the fit. Note that the measurements are centred at the SM prediction for the fit. With linear parametrisations, the impact on the global fit of the entanglement measurements is still very small. Progress in modelling the threshold region and additional data may enhance the weight of this type of measurements in a global analysis in the future.

\begin{figure}[h!]
\includegraphics[width=1.0\columnwidth]{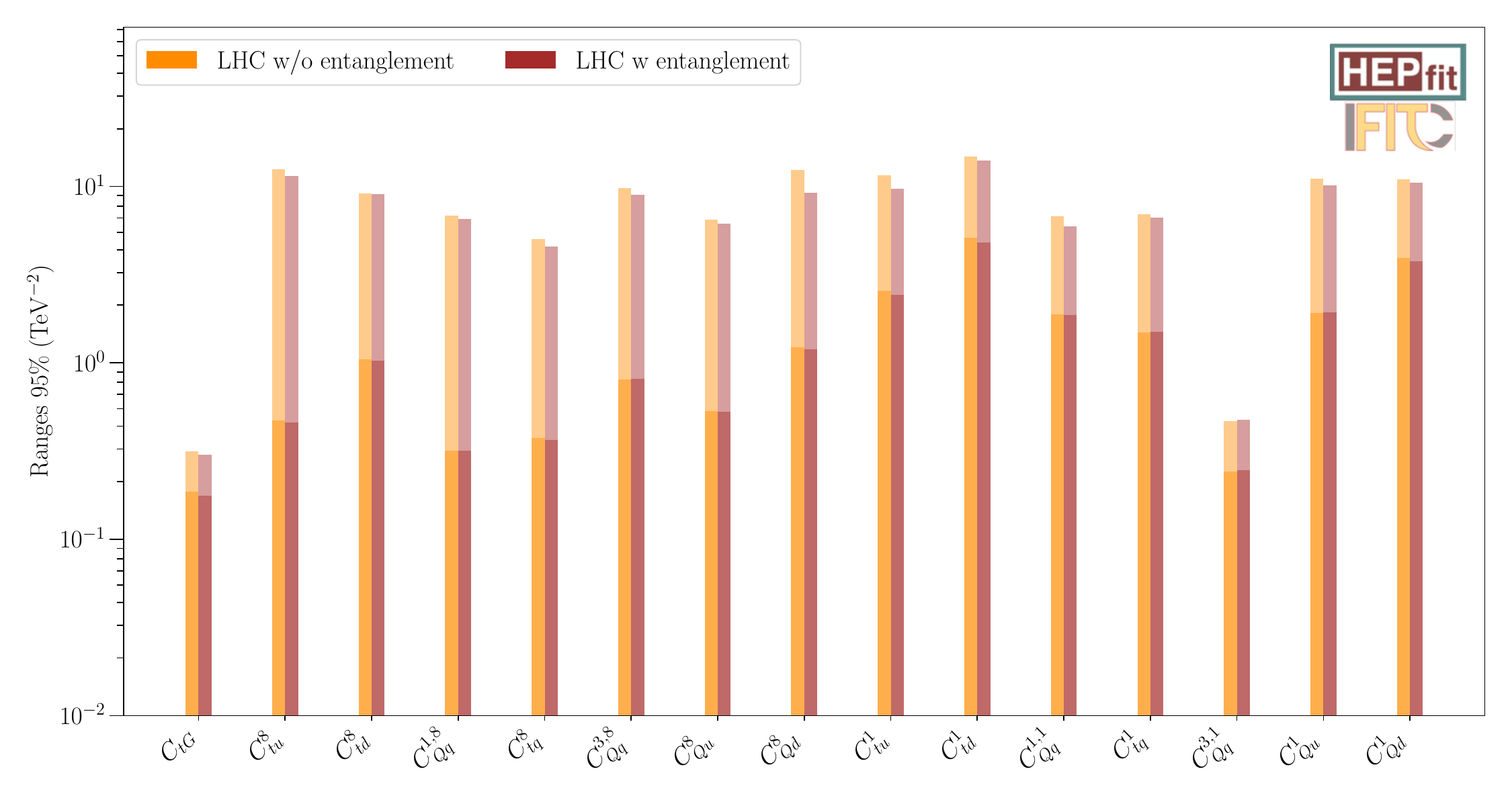}
\caption{\label{fig:entanglement}%
The 95\% probability constraints on the WC divided by $\Lambda^2$ affecting the top-quark entanglement measurement. Solid bars represent individual bounds derived from single-parameter fits, while the shaded regions (full bars) indicate global marginalised constraints obtained by simultaneously fitting all WC.
}
\end{figure}

\section{Prospects for the HL-LHC}
\label{sec:hl-lhc}

The projected precision after completion of the high-luminosity phase of LHC operation is derived from current bounds. A simple extrapolation is employed, based on the ``S2'' scenario in Ref.~\cite{Azzi:2019yne}. This scenario scales statistical and experimental systematic uncertainties with the inverse of the square root of the integrated luminosity. The ``S2'' scenario reduces the theoretical and modelling uncertainties to half of their current magnitude. Note that the measurements are centred at the SM prediction for the fit.

\subsection{Top-quark pair production}

 The LHC has produced an enormous sample of top-quark pairs. After the LHC Run 2, measurements are dominated by hard-to-reduce systematic uncertainties. Therefore, the aggressive scaling of systematic uncertainties assumed in the ``S2'' scenario is replaced by less ambitious projections for this process. This is clearly the case for the inclusive top-quark pair production cross section measurement, that is limited by the jet energy scale and the luminosity calibration~\cite{ATLAS:2023gsl}. The uncertainty on the inclusive cross section is expected to be reduced from 2\% to approximately 1\% at the HL-LHC, a much more modest improvement than what is expected from the ``S2'' scenario.

The comparison with the SM of absolute cross section measurements is limited by the accuracy of the theory. For predictions to catch up with the experimental precision, N$^3$LO corrections are required, along with an important improvement of the proton PDFs. 

The charge asymmetry in top-quark pair production provides complementary information~\cite{Rosello:2015sck}. Here, theory is less of a limiting factor, as the asymmetry can be precisely predicted~\cite{Czakon:2017lgo}. In this case, all experimental systematic uncertainties are improved by a factor 1/2 and only the statistical uncertainty is assumed to scale with the inverse of the integrated luminosity. 

In top-quark pair production, progress is assumed to be driven by a deeper exploration of the boosted regime, rather than by a strong reduction of the uncertainty on the cross section measurement for ``bulk'' top-quark pair production. Therefore,
we have assumed that the last bins of the differential cross section and the charged asymmetry will be further divided by end of the last stage of the HL-LHC, as shown in Appendix C of Ref.~\cite{Durieux:2022cvf}.

\subsection{Rare associated production processes} 

The ``S2'' scenario assumes N$^2$LO calculations will be achieved for rare associated production processes, a milestone that has been reached already for $t\bar{t}H$~\cite{Catani:2022mfv} and $t\bar{t}W$ production~\cite{Buonocore:2023ljm}. In order to reduce modelling uncertainties, Monte Carlo generators for rare processes must be improved in the next decade.
In practice, with these assumptions, theory and modelling are expected to become the dominant uncertainties for most processes, and are the main limiting factor for the SMEFT fit of the top-quark sector at the end of the HL-LHC.

As in top-quark pair production, differential measurements are expected to play an important role in associated production processes. Since the observation of $t\bar{t}X$ processes earlier in the LHC program, analyses have entered the realm of precision physics. Differential measurements are included in our analysis for the $pp \rightarrow t\bar{t}Z$~\cite{ATLAS:2023eld} and $pp \rightarrow t\bar{t}\gamma$ processes~\cite{ATLAS:2024hmk}. The high-$p_{\mathrm{T}}$ bins enhance the sensitivity to the dipole operators~\cite{Bylund:2016phk}, and increase the impact of additional data.

\subsection{Associated production with additional charged leptons}

The analysis considers several degrees of freedom that correspond to operators with two charged leptons and two top quarks, listed in the lowermost box of Tab.~\ref{tab:WilsonOperators}. The LHC experiments can constrain their WC through the study of top-quark production in association with additional charged leptons. 
Bounds of order one are reported from a SMEFT fit to LHC Run~2 data by the CMS experiment~\cite{CMS:2023xyc}. 

The projection of HL-LHC bounds on these operator coefficients is based on an extrapolation of the measurements in $pp \rightarrow t\bar{t}Z$ production~\cite{ATLAS:2021fzm}. The analysis targets events with three or four isolated charged leptons (electrons and muons). Instead of the events with $m_{ll} \sim m_Z$ used in Ref.~\cite{ATLAS:2021fzm}, here we focus on {\em off-shell} events, where the invariant mass of the di-lepton system lies outside the $Z$-boson mass window. Events are binned in four $m_{ll}$ bins: [100-120] \GeV, [120-140] \GeV, [140-180] \GeV and $m_{ll} > $ 180~\GeV~\cite{tesina_abel}. The observed event yields of the Run~2 analysis range from $\sim$~30 events in the lowest bin to about 10 events for the highest bin. 
Individual bounds based on Run~2 data with quadratic parametrisations of $\Lambda^{-2}$ and $\Lambda^{-4}$ terms yield results similar to Ref.~\cite{CMS:2023xyc}, where the sensitivity is driven by the $\Lambda^{-4}$ terms and most pronounced in the highest mass bin.\footnote{Shortly after the release of this manuscript the ATLAS collaboration published a measurement and EFT interpretation of the $t\bar{t}l^+l^-$ process \cite{ATLAS:2025yww}. Our estimates and conclusions are compatible also to those ATLAS results.}

Projections for the HL-LHC are obtained by scaling the integrated luminosity from 140~\ifb{} to 3~\iab. This reduces the statistical uncertainty on the differential cross section to 10-20\%. Systematic uncertainties are expected to remain negligible in comparison to statistical uncertainties.

\subsection{Projections}

HL-LHC projections are compared to existing bounds for the WC of the three classes of operators in Tab.~\ref{tab:WilsonOperators}; the three two-quark two-lepton operators, which we cannot constrain with our dataset without the lepton collider data, are left out of the fit. The results in Fig.~\ref{fig:hllhc_projection_2-quark} correspond to the eight two-fermion operators, the four $e^+e^-t\bar{t}$ operators and those in Fig.~\ref{fig:hllhc_projection_4-quark} to the fourteen $q\bar{q}t\bar{t}$ operators. The projected constraints are presented in terms of the 95\% probability interval of a global fit to all coefficients (full bar). Individual bounds, from fits where the coefficient in question is the only degree of freedom and all other coefficients are set to 0, are indicated with a darker shading. 

\begin{figure}[htp!]
\includegraphics[width=1.0\columnwidth]{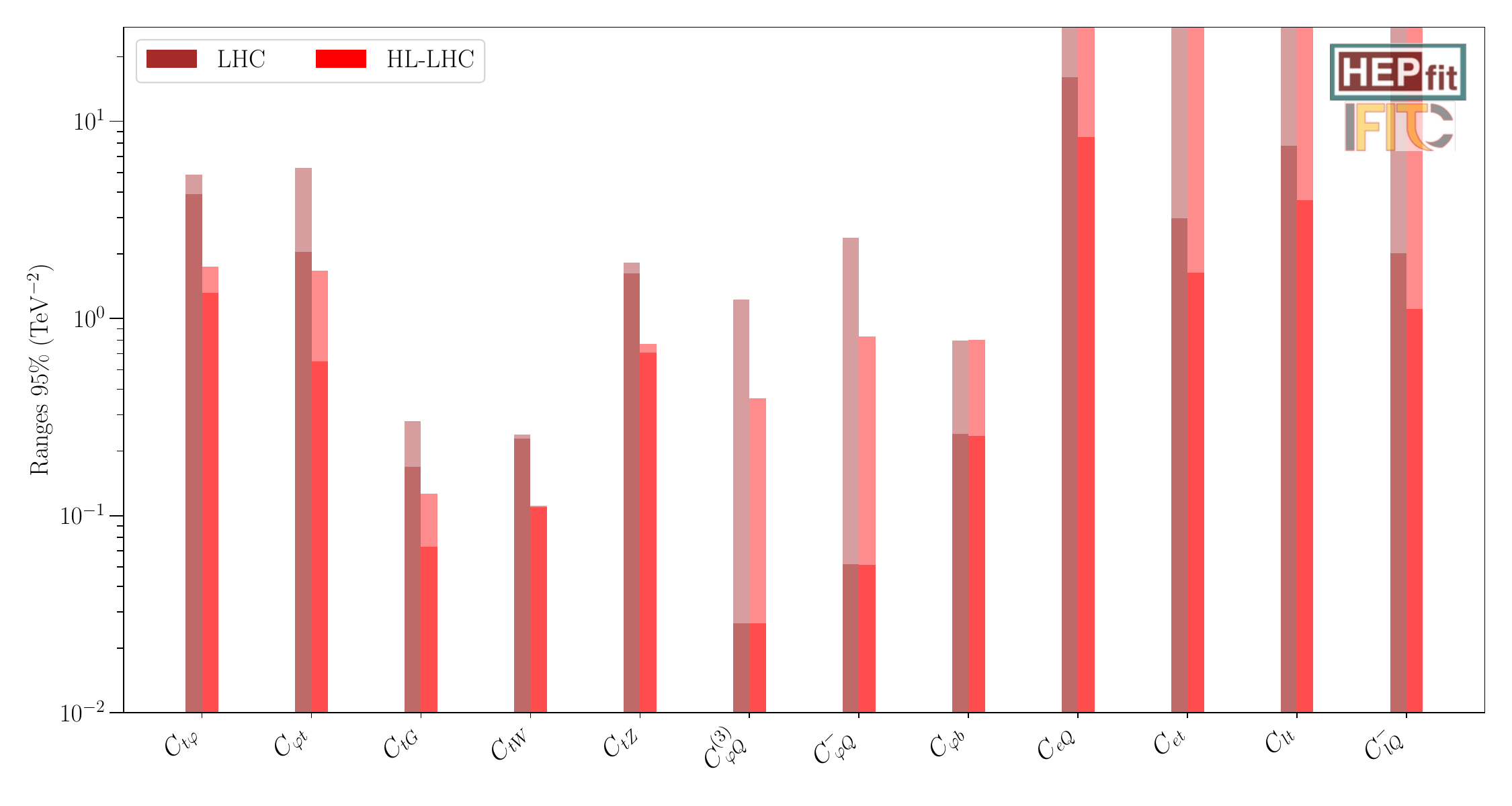}
\caption{\label{fig:hllhc_projection_2-quark}%
The 95\% probability constraints on the WC divided by $\Lambda^2$ for the two-quark dimension-six operators. Both the LHC Run 2 (dark red) and the projections for the HL-LHC (light red) are shown.  
Solid bars represent individual bounds derived from single-parameter fits, while the shaded regions (full bars) indicate global marginalised constraints obtained by simultaneously fitting all Wilson coefficients.
}
\end{figure}

\begin{figure}[htb!]
\includegraphics[width=1.0\columnwidth]{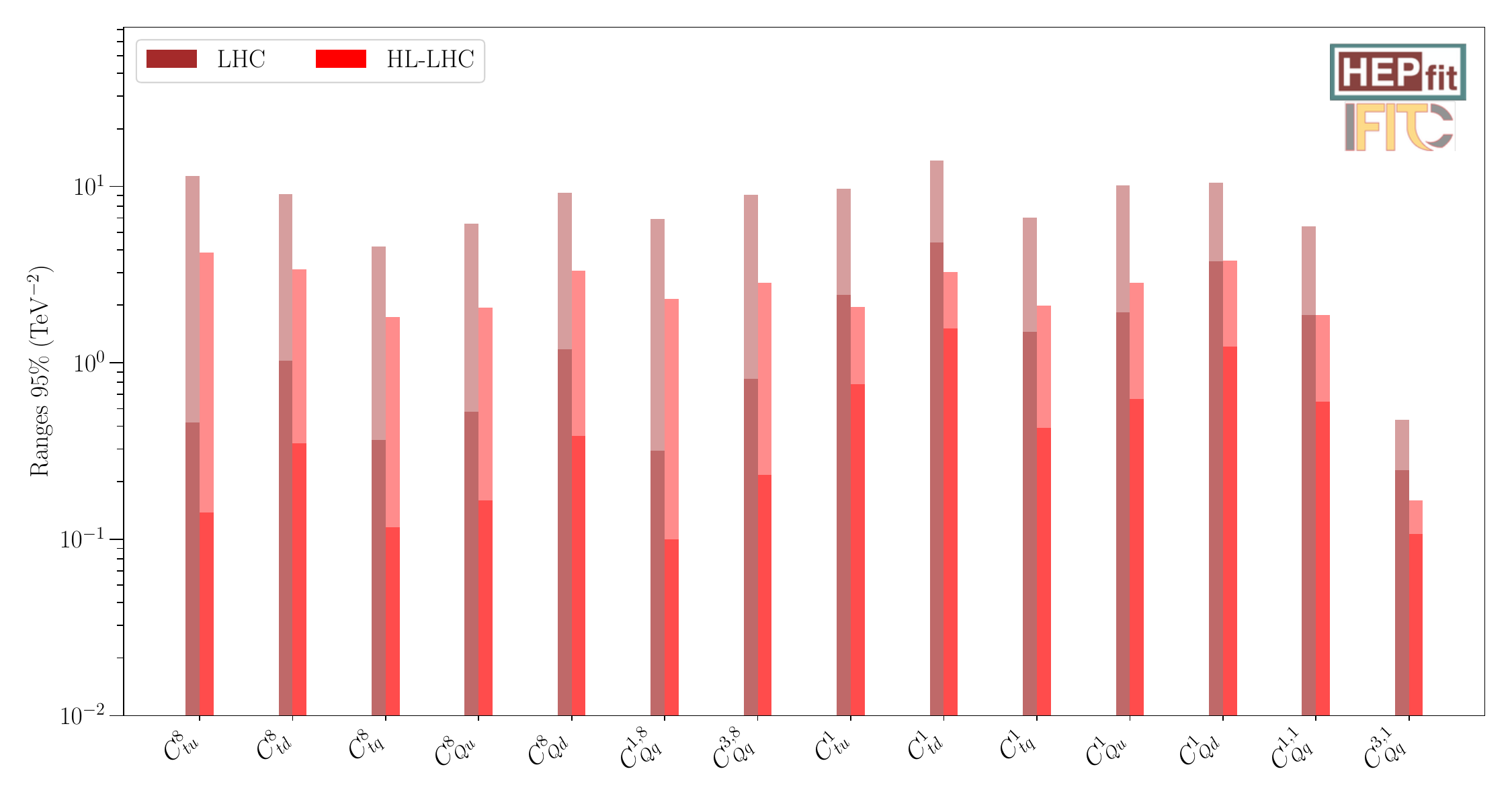}
\caption{\label{fig:hllhc_projection_4-quark}%
Same as Fig.~\ref{fig:hllhc_projection_2-quark} but for the 4-quark operators.}
\end{figure}

The HL-LHC program is expected to improve both the individual and global bounds by a factor of two to four. For two operators, $C_{\varphi Q}^-$ and $C_{\varphi Q}^3$, bounds are dominated by the $Zb\bar{b}$ measurements at the $Z$-pole and HL-LHC brings no progress to the individual bounds. 

In the ``S2'' scenario progress is limited by the accuracy of SM predictions and by modelling uncertainties. Therefore, improving the accuracy of fixed-order predictions beyond the factor two envisaged in the ``S2'' scenario will lead to a direct improvement of the sensitivity. However, this will likely require an accuracy beyond N$^3$LO for $2 \rightarrow 2$ processes and beyond N$^2$LO precision for $2 \rightarrow 3$ processes with top quarks in the final state.

Marginalised bounds remain significantly less precise than individual bounds, even after the complete HL-LHC program, due to ``blind directions'', i.e. unresolved correlations between the large number of degrees of freedom. This is confirmed in fits to the top-quark sector by other groups~\cite{Brivio:2019ius,Hartland:2019bjb} and even in global Higgs/EW/top fits~\cite{Celada:2024mcf,Ethier:2021bye,Ellis:2020unq}. The blind directions have a less pronounced effect in quadratic fits, where dimension-six-squared terms proportional to $\Lambda^{-4}$ are included~\cite{Ethier:2021bye}, as strictly positive contributions cannot cancel each other.

For the two-quark two-lepton operators, the differential measurements of $t\bar{t}l^+l^-$ expected from LHC, or even at HL-LHC, are not enough to lift the degeneracy among the four $t\bar{t}l^+l^-$ operators and provide limits within a reasonable perturbative limit of the operators, even for the scale of NP as low as 1~TeV. Therefore, in order to constrain these operators, the data from a possible future lepton collider are essential, as we will show in next section.

\section{Prospects for \texorpdfstring{$e^+e^-$}{e+e-} colliders}
\label{sec:ee}

The $e^+e^-$ colliders offer an excellent opportunity to study top-quark physics, particularly when operating above the $t\bar{t}$ production threshold. In Tab.~\ref{tab:epem_setup}, we summarise the $e^+e^-$ collider scenarios considered in this work, incorporating the latest specifications from the proposed experiments.

\begin{table}[tb]
    \centering
    \renewcommand{\arraystretch}{1.1}
    \begin{tabular*}{\textwidth}{@{\extracolsep{\fill}}|c|c@{\quad}|c@{\quad}|c@{\quad}|c@{\quad}|}
    \hline
       Machine &  P($e^+$, $e^-$)  & Energy & Luminosity & Reference\\\hline
        \multirow{3}{*}{ILC} &   \multirow{2}{*}{$(\pm30\%,\,\mp80\%)$}  & 250 GeV & 2 \iab{} & \multirow{3}{*}{\cite{AlexanderAryshev:2022pkx}} \\
         &  & 500 GeV & 4 \iab{} & \\
         & $(\pm20\%,\,\mp80\%)$ & 1 TeV & 8 \iab{} & \\\hline
        \multirow{3}{*}{CLIC} & \multirow{3}{*}{$(0\%,\,\pm80\%)$} &  380 GeV & 1 \iab{} & \multirow{3}{*}{\cite{Robson:2018zje}}\\
         &  & 1.5 TeV & 2.5 \iab{} & \\
         &  & 3 TeV & 5 \iab{} & \\\hline
         \multirow{4}{*}{FCC-$ee$} & \multirow{4}{*}{Unpolarised} &  Z-pole & 205 \iab{} & \multirow{4}{*}{\cite{FCC:2025lpp}}\\
         &  & 240 GeV & 10.8 \iab{} & \\
         &  & 350 GeV & 0.42 \iab{} & \\
         &  & 365 GeV & 2.70 \iab{} & \\\hline
          \multirow{4}{*}{CEPC} & \multirow{4}{*}{Unpolarised} &  Z-pole & 57.5 \iab{} & \multirow{4}{*}{\cite{CEPCPhysicsStudyGroup:2022uwl}}\\
         &  & 240 GeV & 20 \iab{} & \\
         &  & 350 GeV & 0.2 \iab{} & \\
         &  & 360 GeV & 1 \iab{} & \\\hline

       \multirow{3}{*}{$\mu$-coll} & \multirow{3}{*}{Unpolarised} &  3 TeV &  1 \iab{} & \multirow{3}{*}{\cite{InternationalMuonCollider:2024jyv}}\\
         &  & 10 TeV & 10 \iab{} & \\
         &  & 30 TeV & 90 \iab{} & \\\hline
    \end{tabular*}
    \caption{Summary of the operating scenarios contemplated for the different lepton collider proposals, indicating the integrated luminosity that is to be collected at different centre-of-mass energies, as well as the beam polarisation. Most numbers remain unchanged with respect to Ref.~\cite{deBlas:2022ofj}, but FCCee has updated its luminosity projection. The luminosity at the ILC is divided among polarisation configurations in the following proportions: $LR:RL:LL:RR=40:40:10:10$. CLIC envisages $LR:RL=50:50$ at $\sqrt{s}=$ 380 GeV and plans to enhance the integrated luminosity in the left-right configuration ($LR:RL=80:20$) at 1.5 TeV and 3 TeV.}
    \label{tab:epem_setup}
\end{table}

\subsection{$e^+e^- \rightarrow b\bar{b}$}

Prospects for bottom-quark pair production are based on the full-simulation studies of the ILD detector concept~\cite{Irles:2024ipg,Irles:2023ojs,Okugawa:2019ycm} at $\sqrt{s} = $ 250~\GeV and 500~\GeV. In these studies, the $b$-tagging efficiency $\epsilon_b$ is determined {\em in-situ} from the measured rates of single and double-tag events. Taking these rates proportional to $2 \epsilon_b (1 - \epsilon_b)$ and $\epsilon^2_b$, respectively, the cross section and the $b$-tagging efficiency can be determined simultaneously~\cite{Irles:2023ojs}. Similarly, a double charge-tag method is used to calibrate the $b/\bar{b}$-tagging. 
In the following, a constant signal acceptance (of 30\% for $R_b$ and 12\% for $A_{FB}$) is assumed, based on the studies of Ref.~\cite{Irles:2024ipg,Irles:2023ojs}. 
These results are extrapolated to higher centre-of-mass energy, assuming the same efficiencies. For the $Z$-pole runs, we use the projections for $R_b$ and $A_{FB}$ provided by the FCCee and CEPC projects for the ``TeraZ'' runs at the $Z$-pole~\cite{deBlas:2022ofj}.

\subsection{$e^+e^- \rightarrow t\bar{t}$}

The top-quark pair production process opens up at $\sqrt{s} \gtrsim$ 350 GeV. It is the dominant six-fermion process and is readily isolated in the fully hadronic, lepton+jets and di-lepton channels~\cite{Amjad:2013tlv,Amjad:2015mma,Durieux:2018tev,CLICdp:2018esa}. This process probes the electroweak couplings of the top quark at tree-level.

The prospects for the top-quark physics program of electron-positron colliders are based on the study of Ref.~\cite{Durieux:2018tev}. The authors define statistically optimal observables in $e^+e^- \to t\bar{t} \rightarrow W^+bW^-\bar{b}$ production~\cite{Durieux:2018tev}, that minimise the hypervolume spanned by the constraints on the relevant WC. The $W^+bW^-\bar{b}$ final state is dominated by top-quark pair production, with a minor contribution from single top-quark production that becomes sizeable only at high centre-of-mass energies. An optimal observable-based analysis of this channel can also be found in Ref.~\cite{Bhattacharya:2023mjr}.

The signal acceptance and identification and reconstruction efficiencies are estimated from full-simulation studies at several centre-of-mass energies in Refs.~\cite{Amjad:2013tlv,Abramowicz:2016zbo}. The efficiency ranges from 10\% close to threshold to 5\% at the highest energy, where the decrease is primarily due to the luminosity spectrum in the high-energy runs at CLIC.\footnote{This effect is much less important in a muon collider, but the efficiency at 3~\tev{} is taken as a {\em conservative} estimate for muon collider operation at 3 TeV, until more complete experimental studies can be done in the challenging environment of a muon collider.} The performance is extrapolated to energies where no full-simulation studies are available.

\subsection{$e^+e^- \rightarrow t\bar{t}H$}\label{sec:eettH}

The golden channel to measure the top-quark Yukawa coupling is the associated $e^+e^- \rightarrow t\bar{t}H$ production process. This process is accessible at centre-of-mass energies above $\sqrt{s}=$ 500--550~\GeV. The projections for the determination of the $t\bar{t}H$ cross section are based on full-simulation studies by ILC and CLIC~\cite{Abramowicz:2016zbo,Price:2014oca,Yonamine:2011jg} and extrapolated where needed.

\subsection{Projections}

\begin{figure}[h!]
\includegraphics[width=1.0\textwidth]{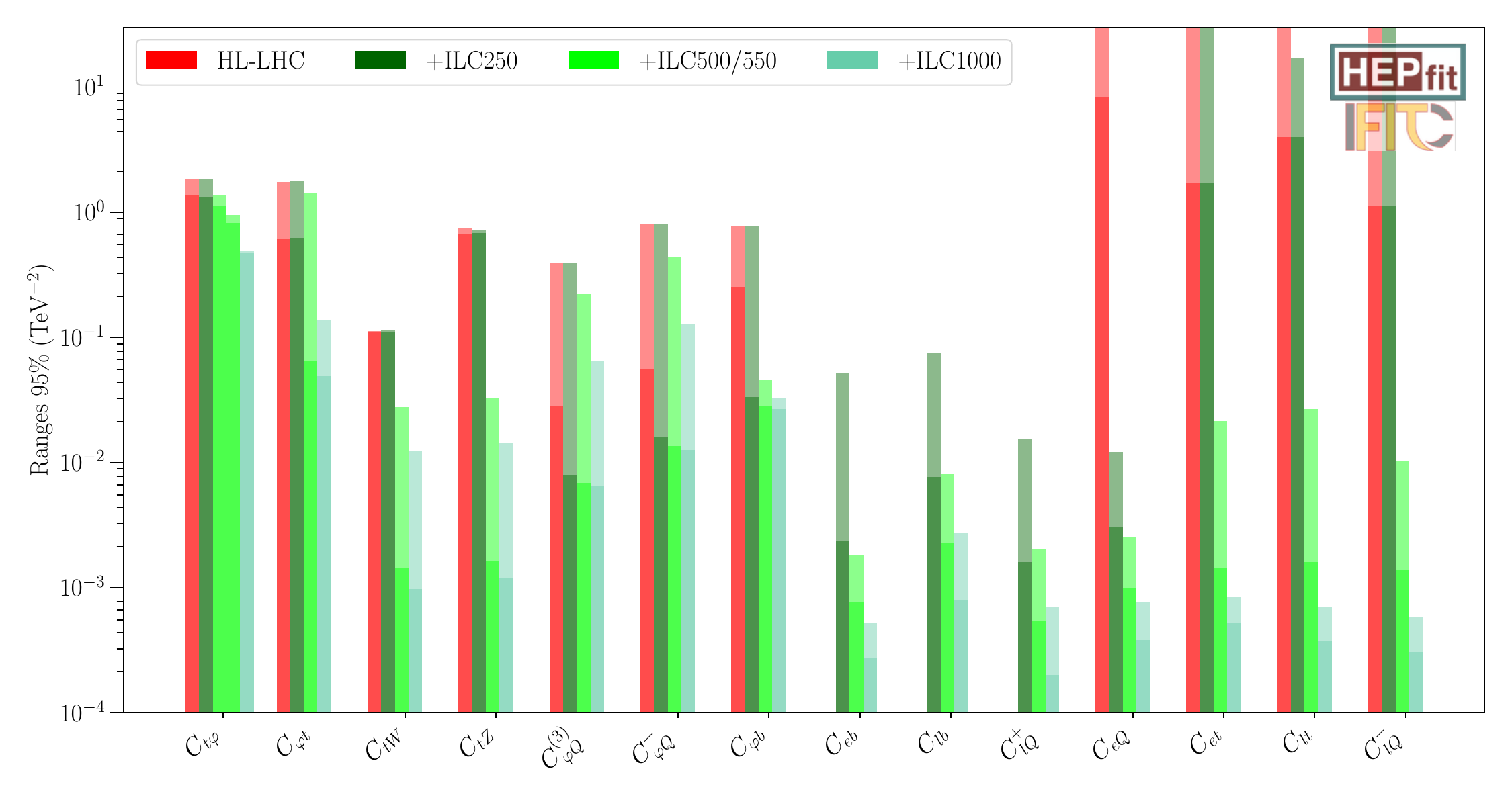}
\caption{\label{fig:hllhc_ILC_projection}%
Comparison of the expected 95\% probability constraints on the WC divided by $\Lambda^2$ at the HL-LHC with those from different stages of the ILC runs at 250, 500 and 1000~GeV.
The limits on the $q\bar{q}t\bar{t}$ and $C_{tG}$ coefficients are not shown, since the $e^+e^-$ collider measurements considered are not sensitive to them, but all operators are included in the global fit.
The improvement expected from the HL-LHC on these coefficients is shown in Figs.~\ref{fig:hllhc_projection_2-quark} and \ref{fig:hllhc_projection_4-quark}. The additional bar included for $C_{t\varphi}$ in light green shows the effect on this operator of ILC working at 550 GeV. The solid bars provide the individual limits of the single-parameter fit and the shaded ones the marginalised limits of the global fit.}
\end{figure}

To assess the potential of electron-positron colliders, the fit is repeated by adding the projected measurements in $e^+e^- \rightarrow b\bar{b}$, $e^+e^- \rightarrow t\bar{t}$ and $e^+e^- \rightarrow t\bar{t}H$ production. These processes are accessible in successive energy stages of the electron-positron collider. The impact of each process and energy stage is made clear for the example of the ILC in Fig.~\ref{fig:hllhc_ILC_projection}.

The first set of red vertical bars shows the HL-LHC projection, with $\mathcal{O}(1\, \TeV^{-2})$ constraints on the two-fermion operators. The global fit finds no meaningful bounds for the $e^+e^-t\bar{t}$ operators $C_{eQ}$, $C_{et}$, $C_{lt}$ and $C_{lQ}^{-}$, or the $e^+e^-b\bar{b}$ operators $C_{eb}$, $C_{lb}$ and $C_{lQ}^{+}$. Note that the $q\bar{q}t\bar{t}$ operator coefficients are not presented in Fig.~\ref{fig:hllhc_ILC_projection}. These receive no new constraints from $e^+e^-$ collisions and their bounds remain at the level of the HL-LHC, given in Fig.~\ref{fig:hllhc_projection_4-quark}.

The first dark green bar adds the ``Higgs factory'' Run at $\sqrt{s}=$ 250~\gev{}. The $e^+e^- \rightarrow b\bar{b}$ measurements provide stringent bounds on the coefficients of the $e^+e^-b\bar{b}$ operators but cannot constrain the top-quark operators. Individual bounds (indicated with darker shading) on the coefficients $C_{\varphi Q}^{(3)}$ and $C_{\varphi}^{-}$ are improved to $\mathcal{O}(10^{-2}\,\TeV^{-2})$ but, since only one linear combination of these two coefficients is constrained, the global bound remains unchanged.     

The second light green bar shows the 95\% probability limits that may be obtained when HL-LHC data are combined with $e^+e^-$ runs at 250~\GeV{} and 500/550~\GeV{}.\footnote{The results for a run at 550 GeV are only shown for the top-quark Yukawa, $C_{t\varphi}$, to show the effect that a slight increase on energy would have on this operator.} Individual bounds for the top-quark dipole operator coefficients $C_{tW}$ and $C_{tZ}$ and the $e^+e^-t\bar{t}$ operator coefficients are improved dramatically. However, global bounds remain relatively weak because of degeneracies among the coefficients. 

The final teal bar shows the impact of adding data at $\sqrt{s}= $ 1~\tev. The sensitivity to four-fermion operators increases with energy, and the probability bounds of 95\% improve by another order of magnitude, reaching values below $10^{-3}~\tev^{-2}$. The individual bounds on two-fermion operators do not improve, but global bounds on these coefficients benefit from the better separation between two-fermion and four-fermion operators. 

The main result of this paper is shown in Fig.~\ref{fig:CEPC_CC_ILC_CLIC_projection}. The plot shows the 95\% probability constraints on the same 14 Wilson coefficients, comparing the operating scenarios of several different electron-positron collider projects. The circular colliders which operate just above the $t\bar{t}$-threshold are not able to constrain the top-quark Yukawa, unlike the linear colliders. Nevertheless, they are able to improve the constraints on several top-quark 2-fermion operators by even some orders of magnitude, especially the individual constraints on top-quark operators affecting the electroweak precision observables, thanks to their ``TeraZ'' runs. They are also to provide constraints on the two-quark two-lepton operators which are not accessible at the HL-LHC. 

\begin{figure}[htb!]
\includegraphics[width=1.0\columnwidth]{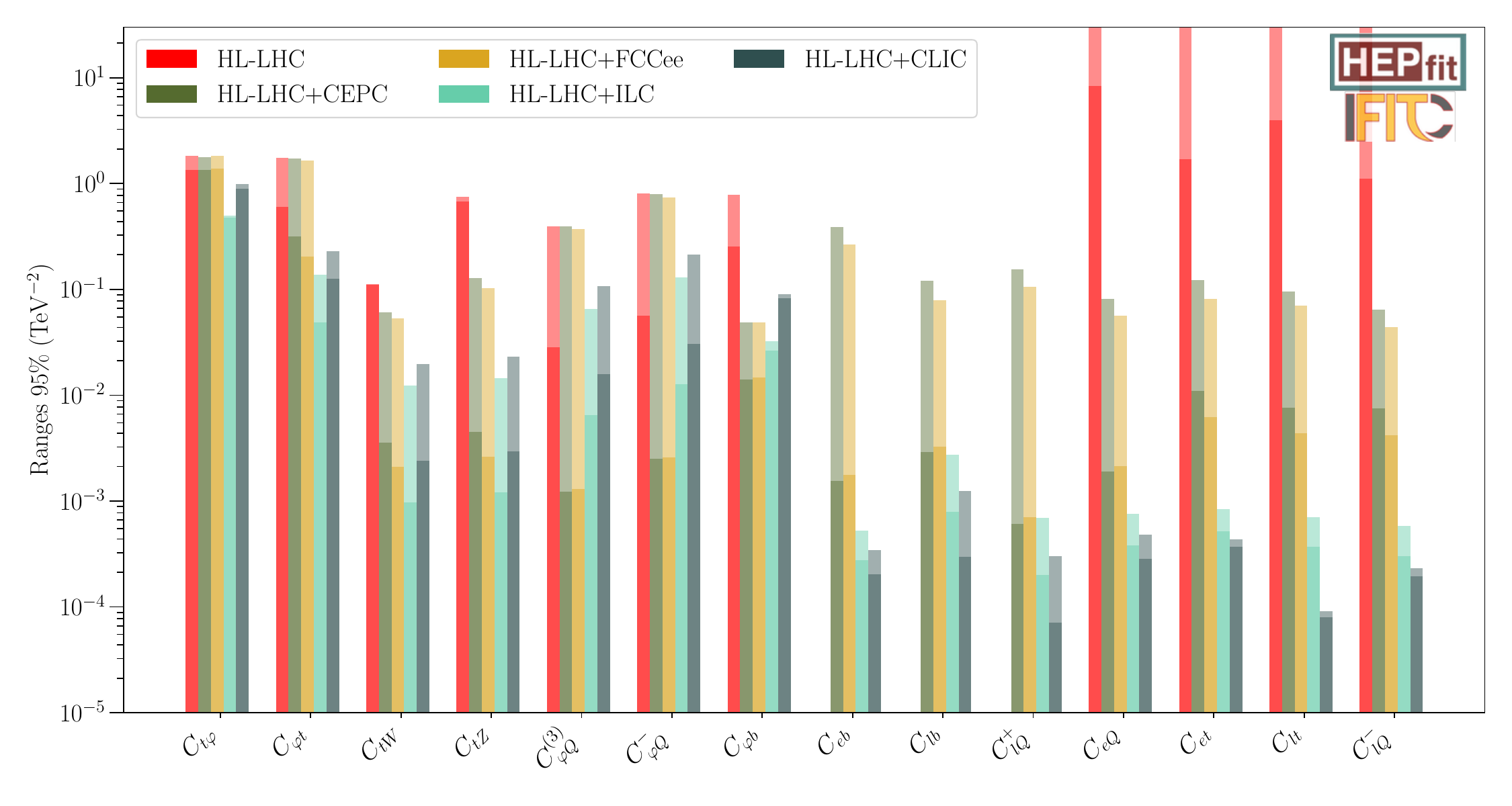}
\caption{\label{fig:CEPC_CC_ILC_CLIC_projection}%
Comparison of the expected 95\% probability constraints on the WC divided by $\Lambda^2$ at the HL-LHC with those from possible future $e^+e^-$ colliders.
The limits on the $q\bar{q}t\bar{t}$ and $C_{tG}$ coefficients are not shown, since the $e^+e^-$ collider measurements considered are not sensitive to them, but all operators are included in the global fit.
The improvement expected from the HL-LHC on these coefficients is shown in Figs.~\ref{fig:hllhc_projection_2-quark}~and~\ref{fig:hllhc_projection_4-quark}. The solid bars provide the individual limits of the single-parameter fit and the shaded ones the marginalised limits of the global fit. 
}
\end{figure}

The linear colliders have several advantages in constraining the top-quark sector including their higher energy reach and the possibility of controlling the polarisation of the initial states. The higher energy span, and especially the separation in the energies above the $t\bar{t}$-thresholds, allows for extremely precise constraints on the two-quark two-lepton operators, as well as on $C_{tW}$ and $C_{tZ}$. The higher energy reach also allows for a significant improvement on the top-quark Yukawa coupling only constrained for energies above 500 GeV in the lepton colliders.

The projections for the top-quark Yukawa, translating $\delta y_t=-\frac{v^2}{\Lambda^2}C_{t\varphi}$, are also presented in Tab.~\ref{tab:top_yuk}. Numbers for lepton colliders are based on an extrapolation in Ref.~\cite{deBlas:2022ofj} of detailed studies in Refs.~\cite{Price:2014oca,CLICdp:2018esa}. We emphasise the effect of increasing the ILC from 500 GeV to 550 GeV at the same integrated luminosity of 4~\iab. The increase in centre-of-mass energy produces a 3.5 factor increase in the cross section which translates into almost a factor of 2 improvement on the individual limit of the Yukawa coupling determination using exclusively the ILC data. Once the HL-LHC data is also considered, the improvement becomes slightly milder, around a 1.4 factor, as shown in Tab.~\ref{tab:top_yuk}.

Higher-energy operation can improve the limit further, as demonstrated by the ILC1000 scenario, with 8~\iab{}. 

\begin{table}[htb!]
    \centering
    \hspace*{-0.5 cm}
    \begin{tabular}{c|c|c|c|c|c|c|c|}
    \multicolumn{2}{c|}{ Uncertainty  }   & LHC  & HL-LHC & ILC500 & ILC550 & ILC1000 & CLIC  \\\hline
    \multirow{4}{*}{$\delta y_t$} & Global  & 16\% & 5.6\%   & 4.1\%   & 2.9\%   & 1.5\%   & 3.0\%  \\
                                  & Individual  & 13\% & 4.1\%   & 3.4\%   & 2.5\%   & 1.5\%   & 2.7\%  \\\cline{2-8}
                                  & Uncombined with & \multirow{2}{*}{--} & \multirow{2}{*}{--}   & \multirow{2}{*}{6.3\%}   & \multirow{2}{*}{3.3\%}   & \multirow{2}{*}{1.6\%}   & \multirow{2}{*}{3.6\%}  \\
                                  &  HL-LHC Indv. &&&&&&
    \end{tabular}
    \caption{
    Uncertainties for the top-quark Yukawa coupling at 68\% probability for different scenarios, in percent. The ILC500, ILC550 and CLIC scenarios also include the HL-LHC, in the top rows. The ILC1000 scenario includes also ILC500 and HL-LHC, for the top rows. Numbers for lepton colliders are based on an extrapolation in Ref.~\cite{deBlas:2022ofj} of detailed studies in Refs.~\cite{Price:2014oca,CLICdp:2018esa}.
    }
    \label{tab:top_yuk}
\end{table}

The results in Tab.~\ref{tab:top_yuk} are generally slightly degraded with respect to earlier studies~\cite{deBlas:2022ofj}, primarily due to the slightly looser bound from the recent LHC result~\cite{ATLAS:2022vkf}, which propagates to the HL-LHC projection. As $e^+e^-$ projections of increasing precision are added, the weight of the HL-LHC result decreases and the result approaches the projections of Ref.~\cite{deBlas:2022ofj}.

The bounds of high-energy colliders can be quite competitive. The inclusion of high-energy muon collider data, reported in detail in the next section, improves the precision of the top-quark Yukawa coupling to 2.5\%. Our projection is less optimistic than the 1.5\% quoted in Ref.~\cite{Liu:2023yrb}, as a result of the more conservative assumptions for the signal acceptance. For reference, the FCChh study of Ref.~\cite{Mangano:2015aow} projects that a precision of 1\% is feasible, provided the pertinent theory improvements. Detailed simulation studies are lacking at this moment, and the control of systematic uncertainties from experimental and theoretical sources remains to be demonstrated.

\section{Lepton colliders in the 3--30~\tev{} range}
\label{sec:muon}

In this section, we study the impact on the top-quark sector of the SMEFT of lepton colliders operated at very high energy, with $\sqrt{s}$ in the range from several to several tens of TeV. This regime could be unlocked by a muon collider, that presents unique benefits (and technological challenges) in comparison with classical electron-positron machines. The Snowmass process provides several compelling reasons to pursue the construction of a Muon Collider \cite{Narain:2022qud}, and the P5 \cite{P5:2023wyd} report has endorsed R$\&$D efforts. Unlike circular electron-positron colliders, muon colliders can reach very high centre-of-mass energies without significant energy losses due to synchrotron radiation. 

At very high energies, a Muon Collider effectively acts as a vector boson collider. The cross section of vector-boson-fusion (VBF) production ($\mu^+\mu^- \to  t\bar{t} \nu \bar{\nu}$ ) increases logarithmically with centre-of-mass energy. At 10-30~\tev{}, the VBF process takes over from the s-channel as the dominant top-quark pair production process. Parametrisations for VBF processes are obtained using \texttt{MadGraph5}\_\texttt{aMC@NLO} with the built-in \textit{muon PDF} that uses the Effective Vector Boson Approximation (EVA)~\cite{Ruiz:2021tdt}. Note that while we use the EVA approximation, the muon PDF without this approximation has been studied in Refs.~\cite{Garosi:2023bvq,Marzocca:2024fqb}.

\subsection{$\mu^+\mu^- \rightarrow b\bar{b}$}

Prospects for bottom-quark pair production on a muon collider, based on full-simulation studies, are not available yet. The extrapolation of the $b$-tagging performance to multi-\TeV{} $b$-jet energy in the challenging environment of a muon collider has considerable uncertainty. Lacking detailed studies, we have extrapolated the constant signal acceptance (of 30\% for $R_b$ and 12\% for $A_{FB}$) from  Ref.~\cite{Irles:2024ipg,Irles:2023ojs} up to the highest energies. This assumption should be refined when full-simulation studies in a realistic environment become feasible.

\subsection{Top-quark pair production in the s-channel : $\mu^+\mu^- \rightarrow t\bar{t}$}

For the case of top-quark pair production, the final state topology depends strongly on centre-of-mass energy. The combined effect of the branching fractions, losses due to the luminosity spectrum and signal selection cuts is taken into account. The fraction of signal events available for use in the analysis is set to 5\% for 3~\tev{} collisions, as in the $e^+e^-$ case~\cite{Durieux:2018tev}. For the 10 TeV case we set the fraction to 2.5\% and at 30 TeV it is reduced to 1\%. This choice may very well prove to be conservative and can be refined as soon as detailed studies are available.

\subsection{Vector-boson-fusion $t\bar{t}$ production: $\mu^+\mu^- \to t \bar{t} \nu \bar{\nu}$ and $t\bar{t} \mu^+ \mu^-$}

At very high energy, vector-boson-fusion production of top-quark pairs becomes important. While the s-channel production falls as $1/s$, the SM cross section for VBF production shows a logarithmic enhancement with centre-of-mass energy. The $\mu^+\mu^- \to t \bar{t} \nu \bar{\nu}$ is dominated by $WW$-fusion diagrams, while $\mu^+\mu^- \to t \bar{t} \mu^+\mu^-$ production has important contributions from $ZZ$, $Z\gamma$ and $\gamma\gamma$ fusion production. The combined cross section grows from 13~fb at $\sqrt{s} = $ 3~\tev{} to 61~fb at $\sqrt{s} = $ 30~\tev. In comparison, the s-channel cross section drops from 19 fb at 3~\TeV to 0.19 fb at 30~\TeV, turning vector-boson-fusion production into the dominant top-quark pair production process above 3~\tev.

The $\mu^+ \mu^- \rightarrow t\bar{t} \mu^+ \mu^-$ and $\mu^+\mu^- \to t \bar{t} \nu \bar{\nu}$ processes have complementary EFT sensitivity. The $W$-boson and $Z$-boson and photon-mediated processes can, at least in principle, be distinguished by tagging the forward muons.  Currently, the instrumentation of the forward region is under development and the practical feasibility of the forward muon tags remains uncertain~\cite{Ruhdorfer:2024dgz}. Therefore, the two processes are analyzed together here.

The sensitivity of the VBF cross section to the Wilson coefficients increases very considerably, in many cases stronger than the SM process. A good example is $C_{ \varphi t}$: the term proportional to $C_{\varphi t} / \Lambda^{-2}$ increases by a factor 10 when the centre-of-mass energy is raised from 3~TeV to 10~TeV, while the quadratic term proportional to $C^2_{t \varphi} / \Lambda^{-4}$ increases by a factor 100. The sensitivity gain in other operator coefficients, such as $C_{t \varphi}$ and the dipole operator coefficients $C_{tW}$ and $C_{tZ}$ is somewhat less pronounced, with factors 3--4 for the linear terms. The term proportional to $C^2_{t \varphi}/ \Lambda^{-4}$, that is very small at low energy, increases by three orders of magnitude for a ten-fold increase of the centre-of-mass energy, as observed in Ref.~\cite{Liu:2023yrb}. With this strong energy growth of the sensitivity, vector-boson-fusion production of top-quark pairs is expected to contribute in an important way to global fits that include the highest energy runs.

\subsection{$\mu^+\mu^- \rightarrow t\bar{t} H$}

As mentioned in section \ref{sec:eettH}, the $t\bar{t}H$ channel provides direct access to constrain the top-quark Yukawa coupling. However, the cross section decreases with energy, from $0.4$ fb at $\sqrt{s}=3$ ~\tev to $0.007$ fb at $\sqrt{s}=30$ ~\tev. On the other hand, VBF-produced $t\bar{t}H$ becomes more significant with increasing energy, surpassing the cross section of the s-channel at around $11.5$ ~\tev. Specifically, the cross section increases from $0.03$ pb at $\sqrt{s}=3$ ~\tev to $0.5$ fb at $\sqrt{s}=30$ ~\tev.

\subsection{Projections}

As discussed in the previous sections, VBF production of top-quark pairs becomes the dominant top-quark pair production process at very high energy. The sensitivity to several operator coefficients moreover grows strongly with energy. Therefore, one expects that at the highest energy the VBF process gains in importance.

The sensitivity of different measurements are compared in Fig.~\ref{fig:muon_collider_sensitivity_twofermion}. The 95\% probability limits are derived from a linear fit measurements of the inclusive $\mu^+\mu^-\to\ttbar$, $\mu^+\mu^- \to \ttbar H$ and VBF production cross sections at three different centre-of-mass energies. We find that VBF bounds improve with centre-of-mass energy, as the cross section, the integrated luminosity and the sensitivity increase with $\sqrt{s}$. Indeed, VBF production at $\sqrt{s}=$ 30~\tev{} provides the best individual bounds on $C_{\varphi t}$ ($\sim$ 0.3~\tev$^{-2}$) and $C_{t \varphi}$ ($\sim$ 4~\tev$^{-2}$), surpassing the s-channel and $t\bar{t}H$ processes. This finding is in qualitative agreement with that of Ref.~\cite{Liu:2023yrb}. The absolute bounds differ mainly due to our more conservative assumptions on the signal acceptance.    
\begin{figure}[htbp!]
    \ 
    \includegraphics[width=1.0\columnwidth]{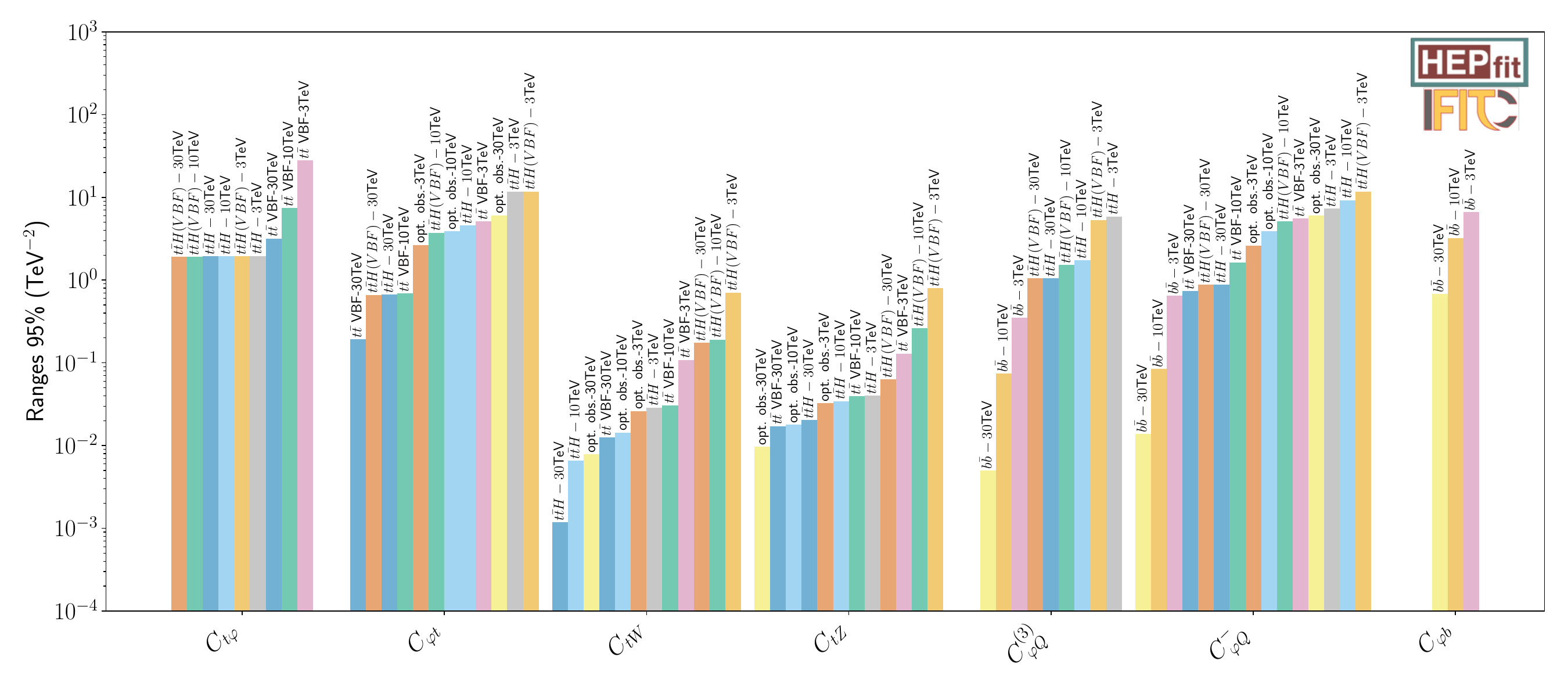} 
    \caption{Comparison of the individual 95\% probability bounds derived from the different measurements. The individual bounds are obtained from fits of a single operator coefficient to a single measurement. }
    \label{fig:muon_collider_sensitivity_twofermion}
\end{figure}

\begin{figure}[h!]
\includegraphics[width=1.0\columnwidth]{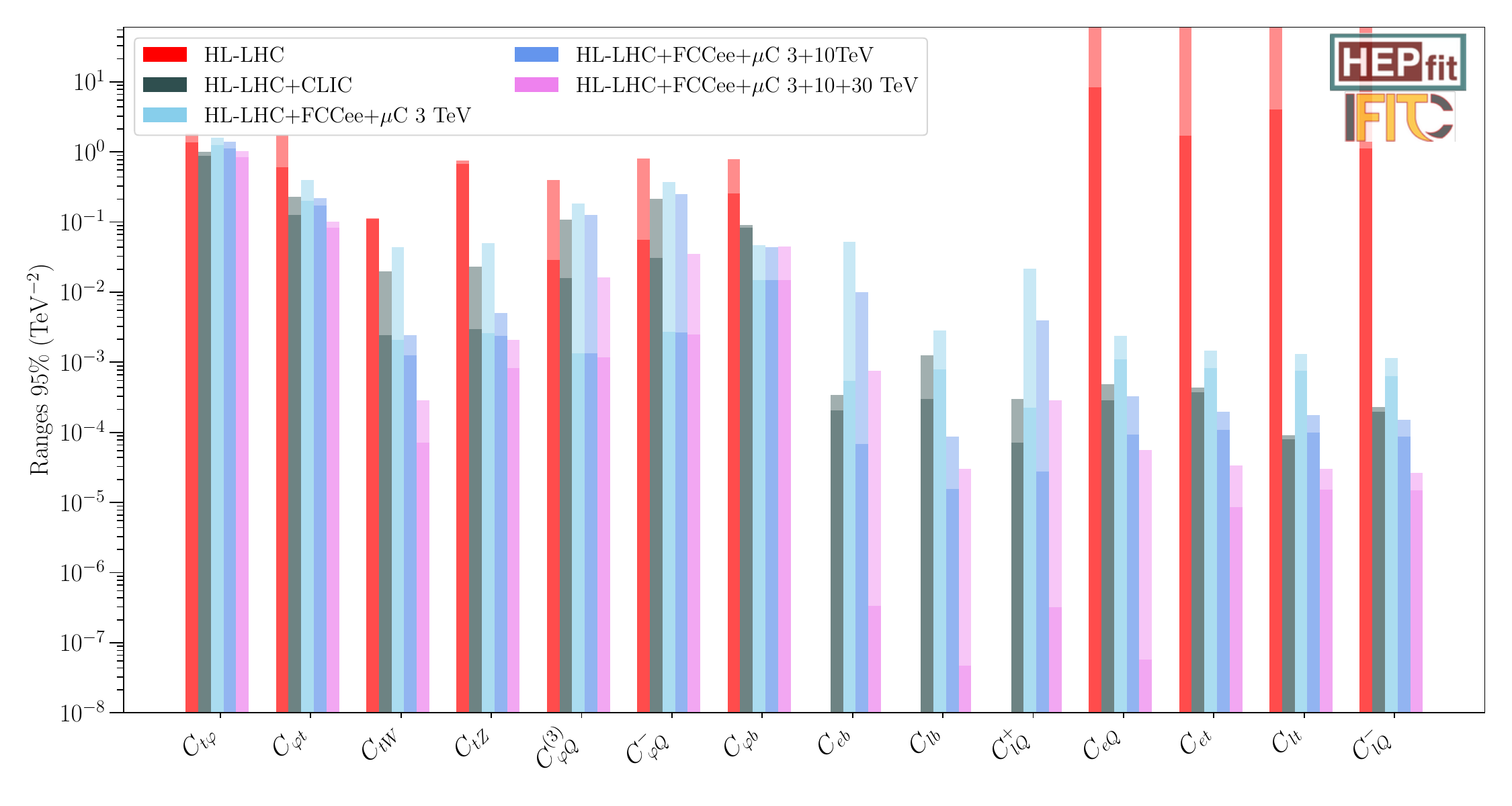}
\caption{\label{fig:muon_projection}%
Comparison of the expected 95\% probability constraints on the WC divided by $\Lambda^2$ at the HL-LHC with those from CLIC and the muon collider.
The limits on the $q\bar{q}t\bar{t}$ and $C_{tG}$ coefficients are not shown, since the lepton collider measurements considered are not sensitive to them, but all operators are included in the global fit.
The improvement expected from the HL-LHC on these coefficients is shown in Figs.~\ref{fig:hllhc_projection_2-quark}~and~\ref{fig:hllhc_projection_4-quark}. The solid bars provide the individual limits of the single-parameter fit and the shaded ones the marginalised limits of the global fit. }
\end{figure}

In Fig.~\ref{fig:muon_projection} we show the projections for the muon collider, comparing them with the HL-LHC and the CLIC scenarios. An interesting feature is that the performance of CLIC at 3 TeV is similar to FCCee and a muon collier at 3 TeV and 10 TeV. Indeed, the possibility of polarised beams allows CLIC to outperform a muon collider at the same energy and it is competitive (or even better) than a muon collider at a much higher energy of 10 TeV.

\section{Conclusion}
\label{sec:conclusion}

In this paper, we have presented a global analysis of the top- and bottom-quark sector of the SMEFT. In our fit, the sensitivity is parametrised considering only {\em linear} terms from the interference between the Standard Model and the dimension-six operators, which are proportional to $\Lambda^{-2}$. 
Compared to previous studies~\cite{Durieux:2019rbz}, we have updated several measurements with recent results based on Run 2 of the LHC. New differential measurements are included in $t\bar{t} H$ production~\cite{ATLAS:2022vkf}, $t\bar{t}Z$ production~\cite{ATLAS:2023eld} and $t\bar{t}\gamma$ production~\cite{ATLAS:2024hmk}, as well as new measurements of the inclusive $t\bar{t}W$, $tZq$ and $t\gamma q$ rates. We include differential measurements of the cross section~\cite{CMS:2021vhb} and charge asymmetry~\cite{ATLAS:2022waa} in top-quark pair production as a function of the invariant mass of the \ttbar{} system. In addition, two measurements sensitive to quantum entanglement~\cite{CMS:2024pts,CMS:2024zkc} from CMS have been included.
Although the LHC Run 2 data clearly dominate, several legacy measurements, such as the forward-backward asymmetry in top-quark pair production and the s-channel single top-quark production cross section at the Tevatron, and the $Z \rightarrow b\bar{b}$ measurements at LEP and SLC remain important.

A fit is performed to 22 Wilson coefficients, corresponding to eight two-fermion operators and fourteen $q\bar{q}t\bar{t}$ operators, to measurements at the LHC, Tevatron, LEP and SLC. This fit yields 95\% probability global bounds ranging from $\cal{O}$(0.1) \tev$^{-2}$ for some two-fermion operators to $\cal{O}$(10) \tev$^{-2}$ for the $q\bar{q}t\bar{t}$ operators. The bounds on the coefficients $C_{\varphi t}$ and $C_{t \varphi}$ that modify the top-quark right-handed coupling and the Yukawa coupling, respectively, remain relatively weak. Individual bounds on four-quark operators are an order of magnitude better than global bounds, due to unresolved correlations between coefficients. The spin correlation measurements that probe quantum entanglement at the \ttbar{} threshold and in the boosted regime are found to have good sensitivity to $C_{tG}$ and the four-quark operator coefficients, as suggested in Ref.~\cite{Severi:2022qjy}, but do not lead to significantly better global bounds at their current precision. 

We also provide projections for the high-luminosity phase of the Large Hadron Collider and future colliders. HL-LHC projections are obtained adopting the ``S2'' scenario that is also used for Higgs measurements~\cite{Cepeda:2019klc}.
In this scenario, experimental uncertainties are strongly reduced $(\propto 1/\sqrt{L_\text{int}})$, while theory and modelling uncertainties see a more modest improvement (by a factor 1/2).
The bounds improve by a factor 2-3 for most of the operator coefficients. 
In the projections, we extend previous studies~\cite{Durieux:2019rbz} by considering also seven operators with two charged leptons and two top quarks, enlarging the total number of Wilson coefficients to 29. In principle, these can be constrained by measurements of $t\bar{t}l^+l^-$ production with $m_{ll} > m_Z$. However, we find that no meaningful global bounds are obtained with a parametrisation based on terms proportional to $\Lambda^{-2}$ .

A lepton collider operated above the $\ttbar$ production threshold has excellent potential to constrain the coefficients of bottom- and top-quark operators. The projections used in this paper are based on full-simulation study of optimal observables in Ref.~\cite{Durieux:2018tev}, with up-to-date operating scenarios for electron-positron Higgs/top/EW factories. Individual bounds on the coefficients $C_{tZ}$ and $C_{tW}$ of the dipole operators reach and on $C_{\varphi Q}^{(1)}$ and $C_{\varphi Q}^{(3)}$ reach 10$^{-2}$ \tev$^{-2}$ with a single run with an integrated luminosity of 4~\iab{} at $\sqrt{s}=$ 500~\gev. The bounds on the two-lepton-two-heavy-quark operators even reach 10$^{-3}$ \TeV$^{-2}$. The global bounds are limited by degeneracies between the two-fermion and four-fermion operators, that can be lifted to a good extent by including measurements at two different centre-of-mass energies. Constraints on the operators considered in our analysis from a possible future lepton collider have also been recently derived in Refs. \cite{Maura:2025rcv,Hoeve:2025yup}.

Lepton collisions at even higher energy, with a centre-of-mass energy up to several tens of TeV, can be achieved in the future with novel acceleration techniques. Concrete proposals have been put forward for a muon collider~\cite{InternationalMuonCollider:2024jyv} and a collider based on high-gradient wakefield acceleration~\cite{Proceedings:2024ncv,ALEGRO:2019alc}. At centre-of-mass energies beyond 3~TeV, vector-boson-fusion production of top-quark pairs becomes the dominant production mechanism and offers competitive bounds on two-fermion operator coefficients, including coefficients such as $C_{t \varphi}$ and $C_{\varphi t}$ that are hard to constrain elsewhere. At the same time, the energy-growth in the sensitivity to two-quark two-lepton operators yields bounds on $l^+l^-t\bar{t}$ coefficients below $10^{-4}$ TeV$^{-2}$ if 10 ab$^{-1}$ is collected at $\sqrt{s} = $ 10~TeV or beyond. Beam polarisation remains an important asset at high energy, with an impact on the bounds of all coefficients. 

\FloatBarrier

\section*{Acknowledgments}
We thank J. Tian and M. Peskin for their comments and suggestions on the first version of the manuscript. The authors acknowledge the work of Master students Abel Gutierrez Camacho, Belén Durán González and Pablo Copete Garrido in the development of this analysis.
F.C.G\ is supported by the Presidential Society of STEM Postdoctoral Fellowship at Case Western Reserve University and by the \textit{Ministerio de Ciencia, Innovaci\'on y Universidades}, Spain, through a Beatriz Galindo Junior grant BG23/00061.
The work of VM has been supported by the Italian Ministry of Research (MUR) under the grant PRIN20172LNEEZ, by the European Research Council (ERC) under the European Union’s Horizon 2020 research and innovation programme (Grant agreement No. 949451) and a Royal Society University Research Fellowship through grant URF/R1/201553. 
The work of MML is supported by the Science and Technology Facilities Council [grant number ST/X005941/1]. 
MMLL acknowledges the support received from the Spanish Ram\'on y Cajal programme (RYC2019-028510-I). Her work is also supported by the projects ASFAE/2022/010 and CIPROM/2022/70 (Generalitat Valenciana), and PID2021-124912NB-I00 (Spanish Ministry MICIN). 
The work of MV is supported by the Spanish ministry for science under grant number PID2021-122134NB-C21, by the Generalitat Valenciana under PROMETEO grant CIPROM/2021-073, and by CSIC under grant ILINKB20065. The group members of IFIC in Valencia received support from the Severo Ochoa excellence programme.

\appendix

\FloatBarrier

\input{appendix_output_correlation_matrices}

\FloatBarrier

\bibliographystyle{apsrev4-1_title}
\bibliography{top.bib}

\end{document}

%% file: appendix_output_correlation_matrices.tex
\section{Appendix: Correlation matrices}
\label{app:correlation_matrices}

In the following, we show the correlation matrices obtained for the different scenarios that we have considered. For all of these results, the experimental values for the observables have been set to the best SM prediction.

\begin{figure}\centering
\includegraphics[width=1.1\columnwidth]{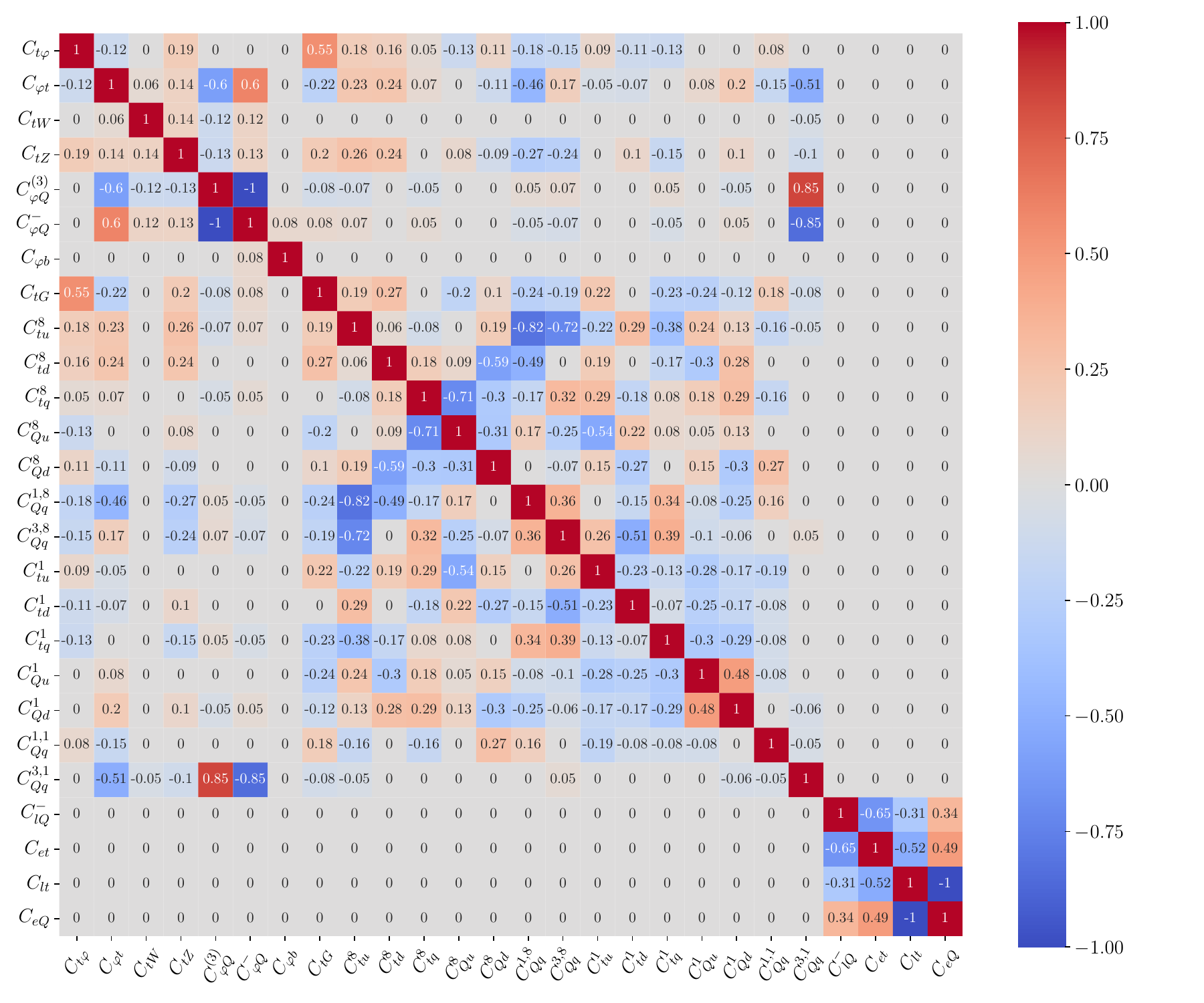}
\caption{\label{fig:corr_LHC_1}Correlation matrix obtained for the global fit including the data of the LHC, Tevatron and LEP.  Entries smaller than 5\% are set to zero.  }
\end{figure}

\begin{figure}\centering
\includegraphics[width=1.1\columnwidth]{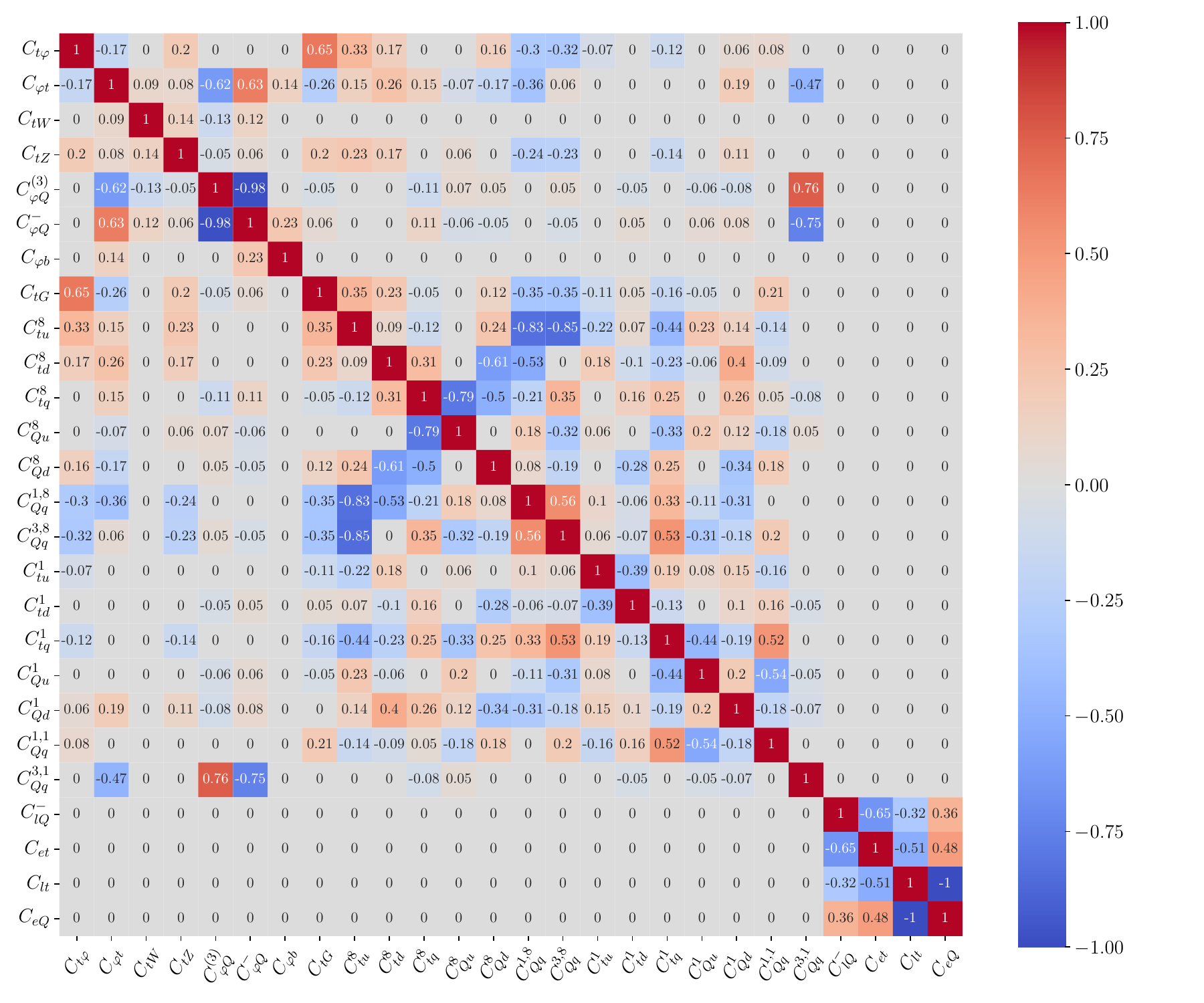}
\caption{\label{fig:corr_LHC_2}Correlation matrix obtained for the global fit including the data of the HL-LHC, Tevatron and LEP.  Entries smaller than 5\% are set to zero.  }
\end{figure}

\begin{figure}\centering
\includegraphics[width=1.1\columnwidth]{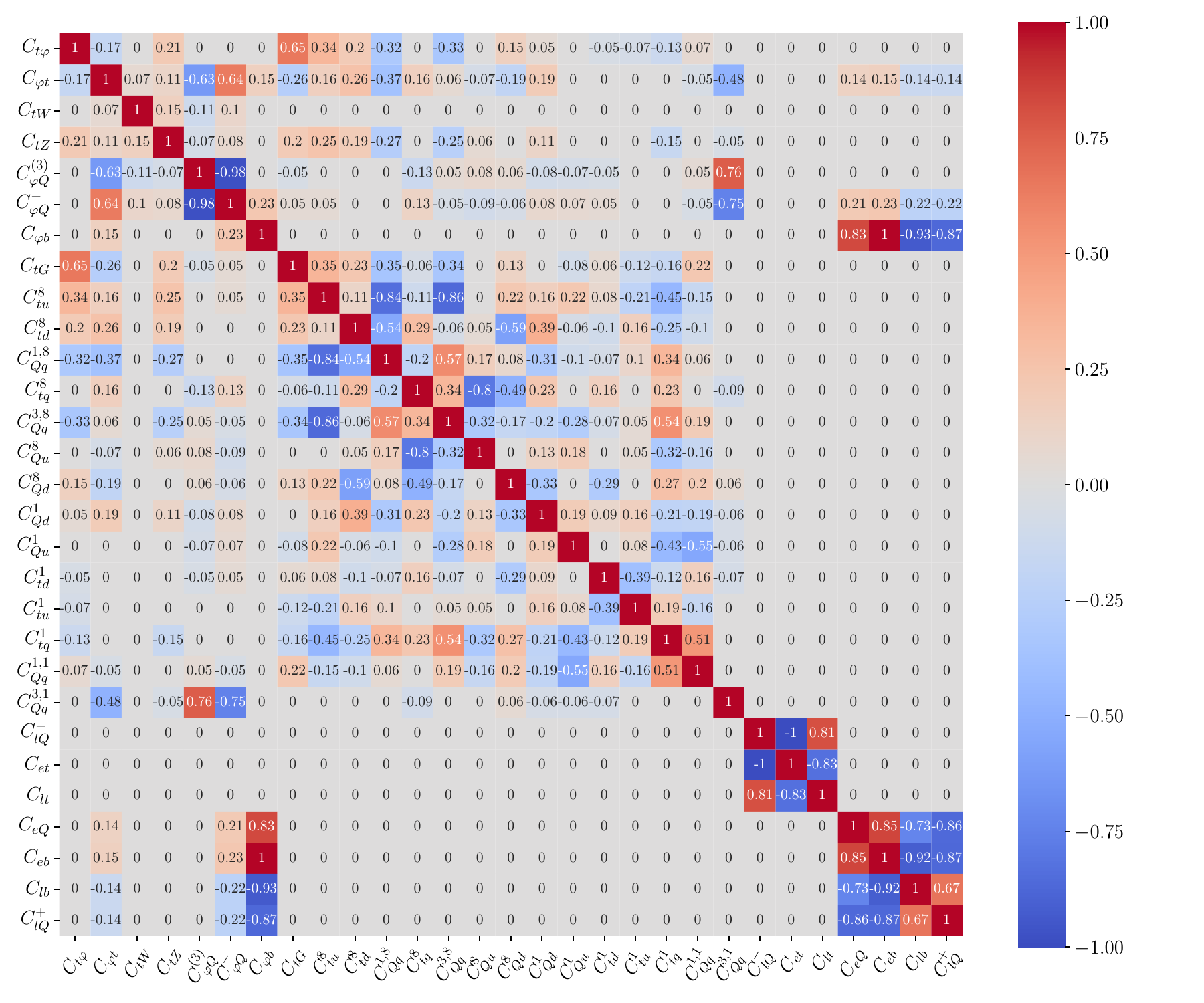}
\caption{\label{fig:corr_LHC_3}Correlation matrix obtained for the global fit including the data of the HL-LHC, Tevatron, LEP and ILC working at 250 GeV.  Entries smaller than 5\% are set to zero. }
\end{figure}

\begin{figure}\centering
\includegraphics[width=1.1\columnwidth]{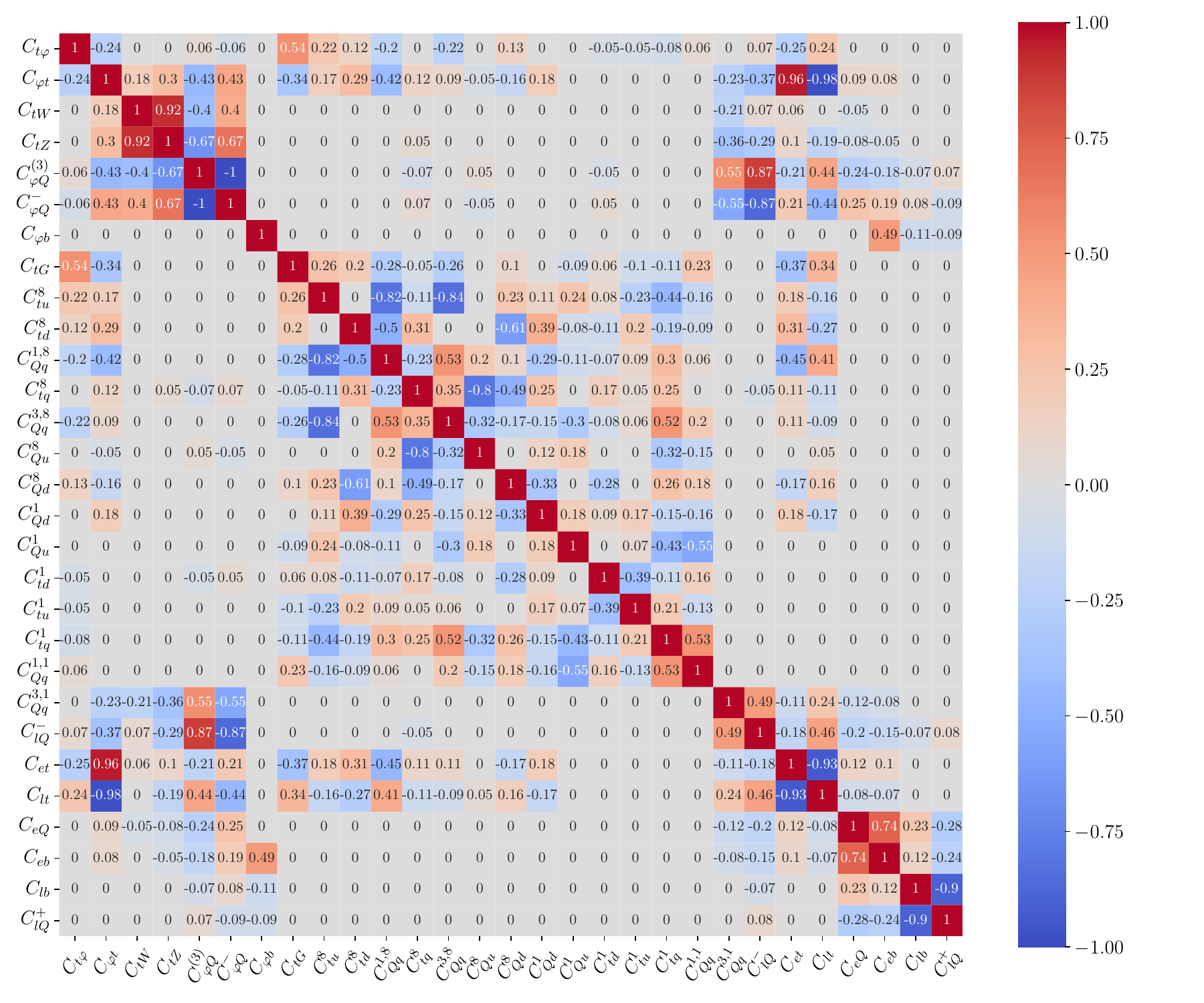}
\caption{\label{fig:corr_LHC_4}Correlation matrix obtained for the global fit including the data of the HL-LHC, Tevatron, LEP and ILC working at 250 GeV and 500 GeV.  Entries smaller than 5\% are set to zero. }
\end{figure}

\begin{figure}\centering
\includegraphics[width=1.1\columnwidth]{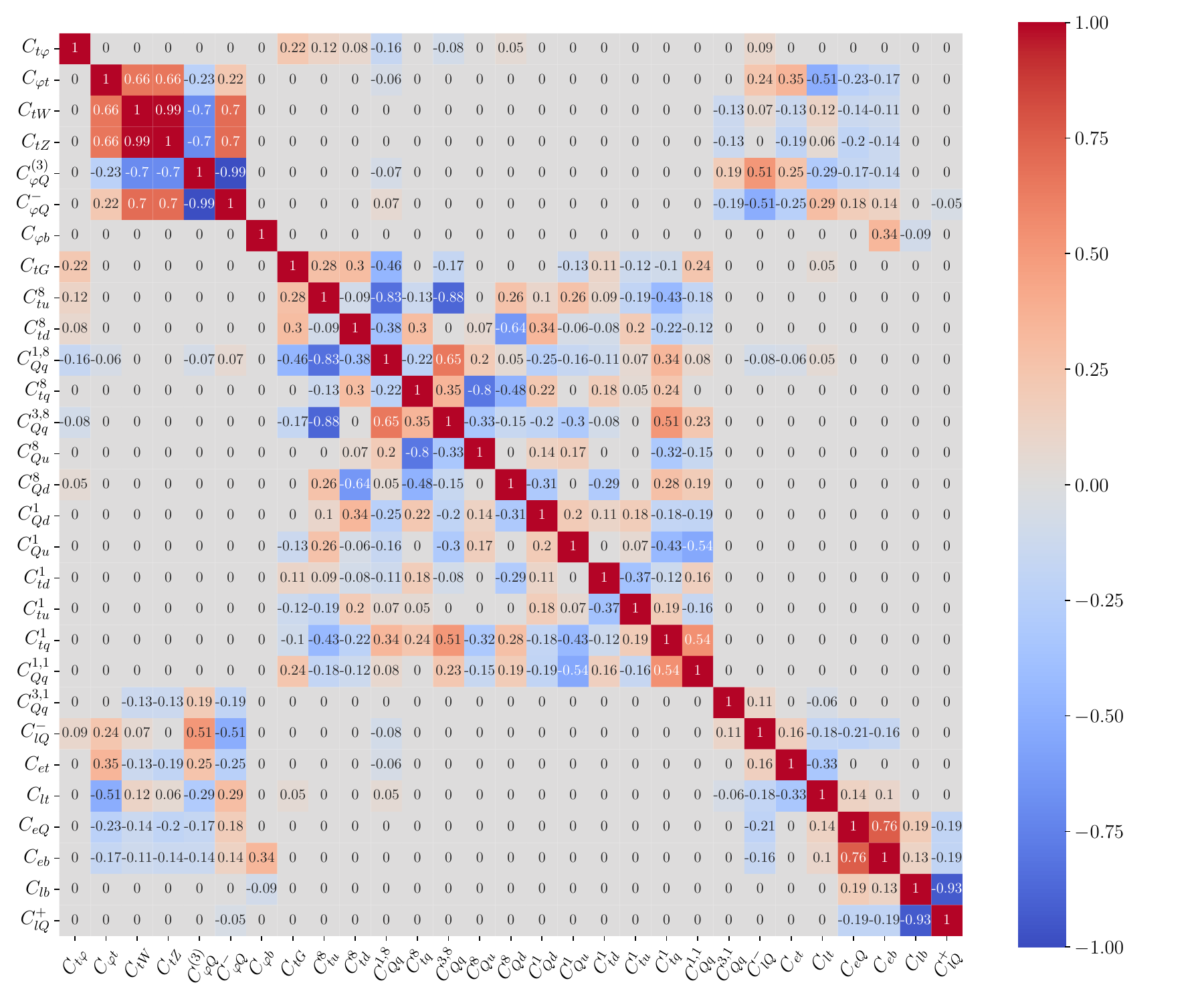}
\caption{\label{fig:corr_LHC_5}Correlation matrix obtained for the global fit including the data of the HL-LHC, Tevatron, LEP and ILC working at 250 GeV, 500 GeV and 1000 GeV.  Entries smaller than 5\% are set to zero. }
\end{figure}

\begin{figure}\centering
\includegraphics[width=1.1\columnwidth]{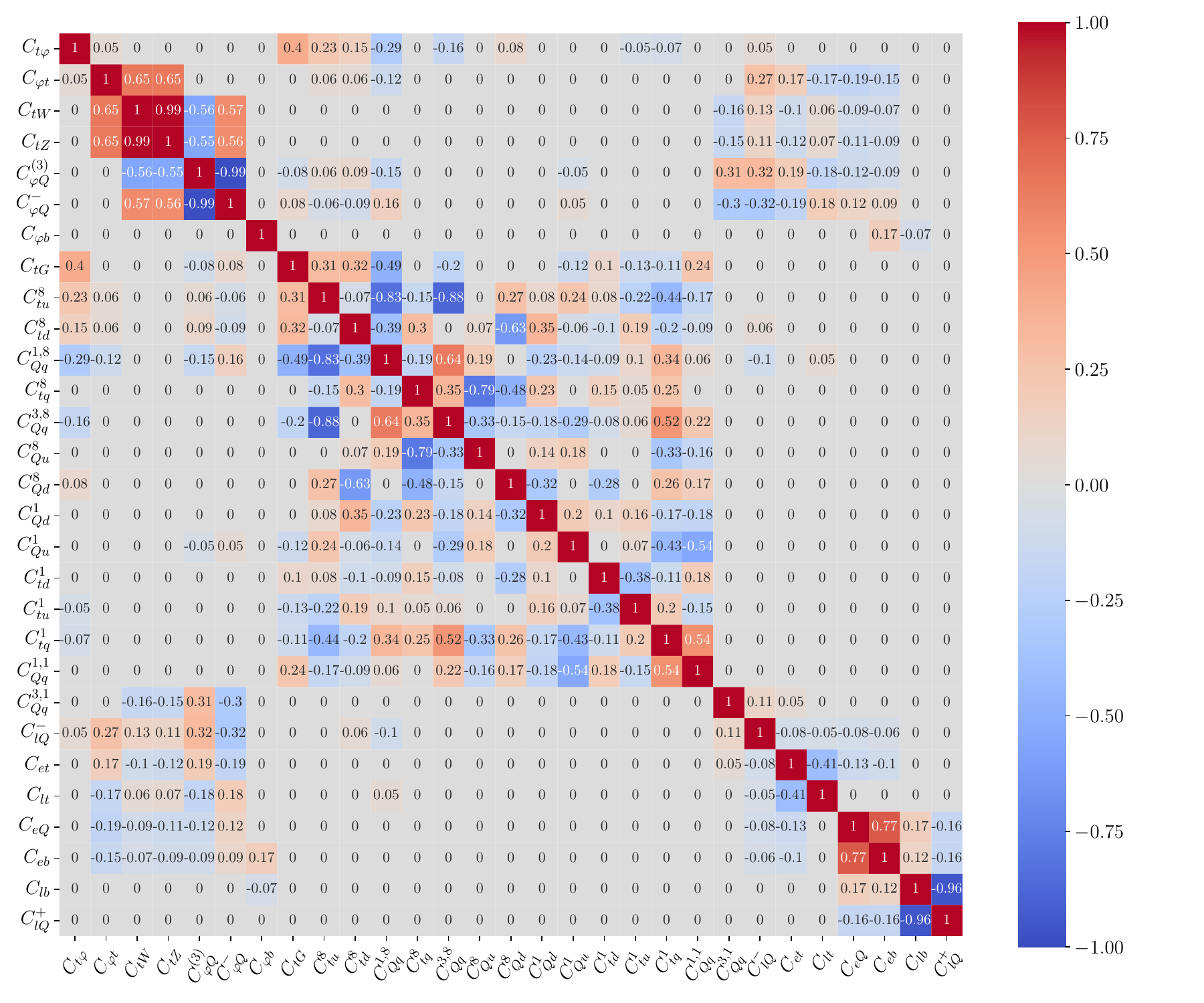}
\caption{\label{fig:corr_LHC_6}Correlation matrix obtained for the global fit including the data of the HL-LHC, Tevatron, LEP and the final stage of CLIC.  Entries smaller than 5\% are set to zero. }
\end{figure}

\begin{figure}\centering
\includegraphics[width=1.1\columnwidth]{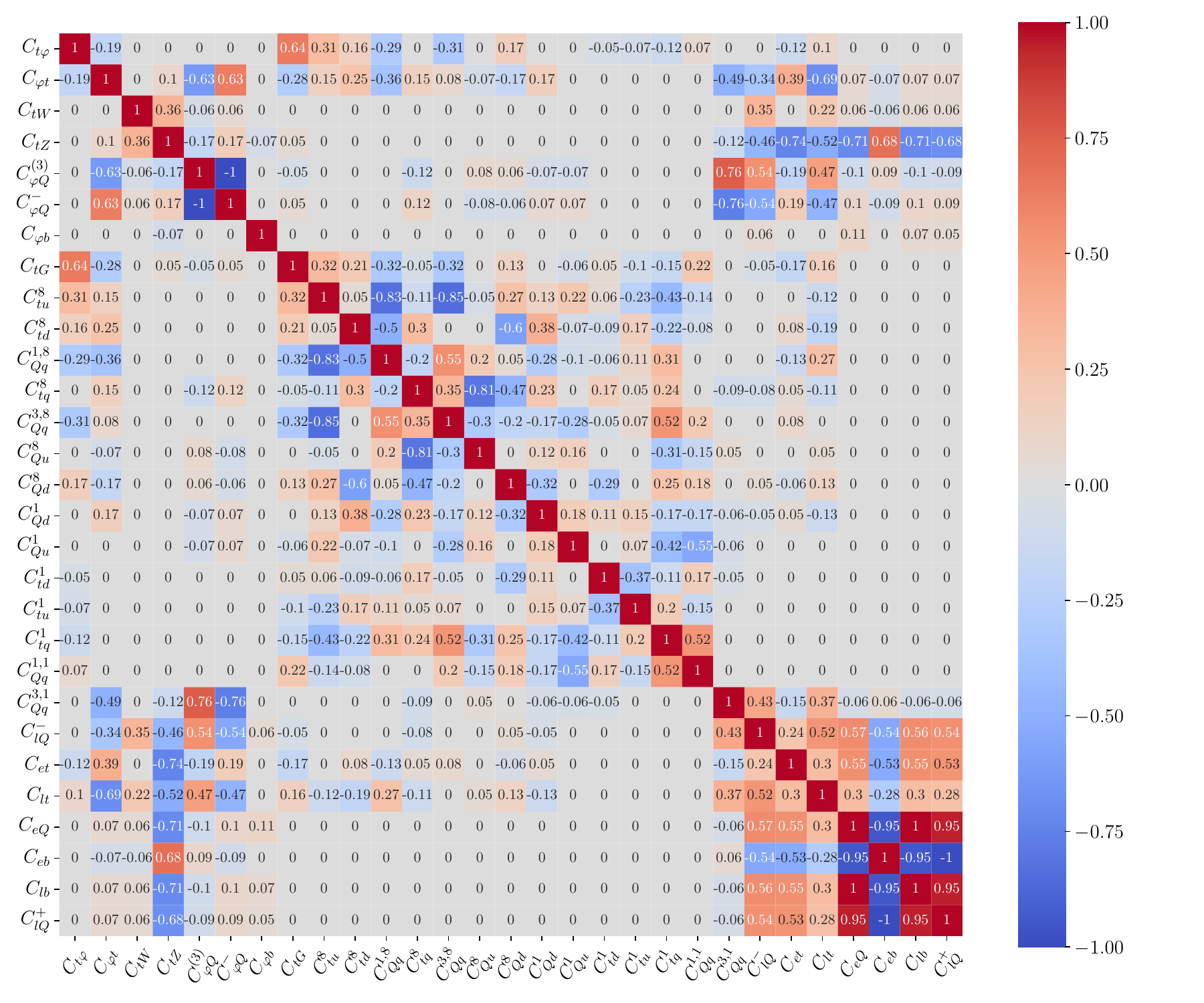}
\caption{\label{fig:corr_LHC_7}Correlation matrix obtained for the global fit including the data of the HL-LHC, Tevatron, LEP and the final stage of CEPC.  Entries smaller than 5\% are set to zero. }
\end{figure}

\begin{figure}\centering
\includegraphics[width=1.1\columnwidth]{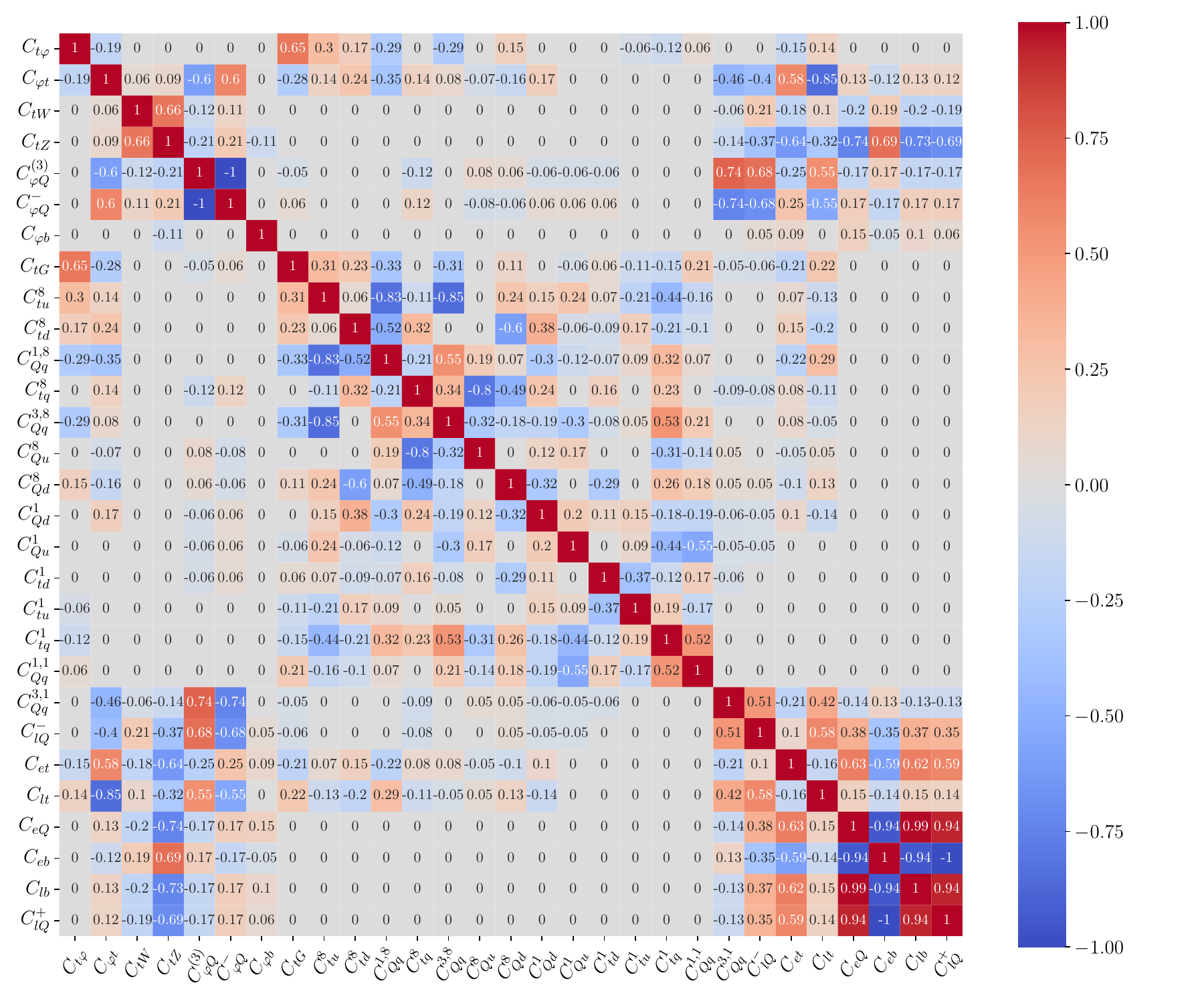}
\caption{\label{fig:corr_LHC_8}Correlation matrix obtained for the global fit including the data of the HL-LHC, Tevatron, LEP and the final stage of FCCee.  Entries smaller than 5\% are set to zero. }
\end{figure}

\begin{figure}\centering
\includegraphics[width=1.1\columnwidth]{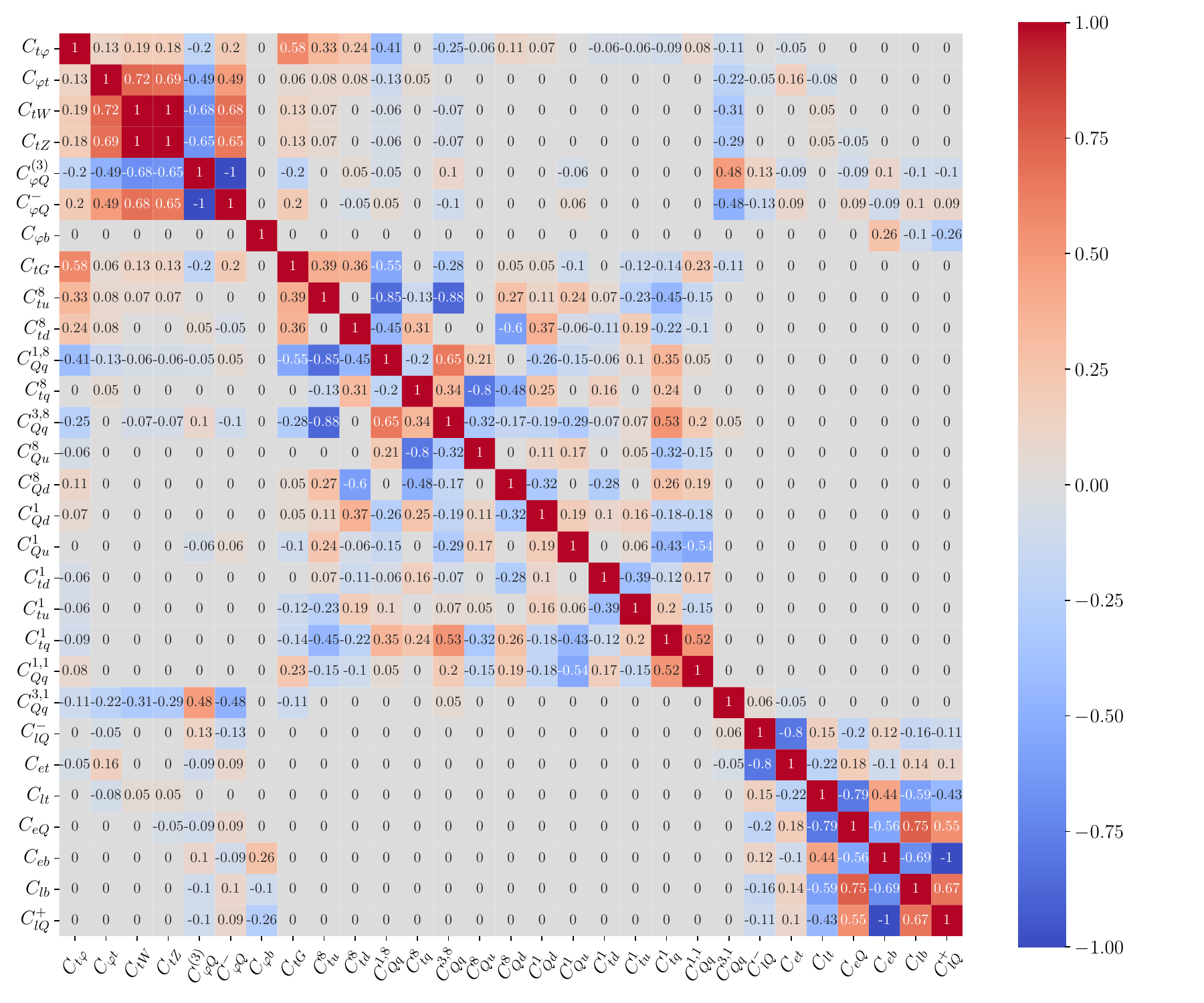}
\caption{\label{fig:corr_LHC_9}Correlation matrix obtained for the global fit including the data of the HL-LHC, Tevatron, LEP, the final stage of FCCee and a muon collider at 3 TeV.  Entries smaller than 5\% are set to zero. }
\end{figure}

\begin{figure}\centering
\includegraphics[width=1.1\columnwidth]{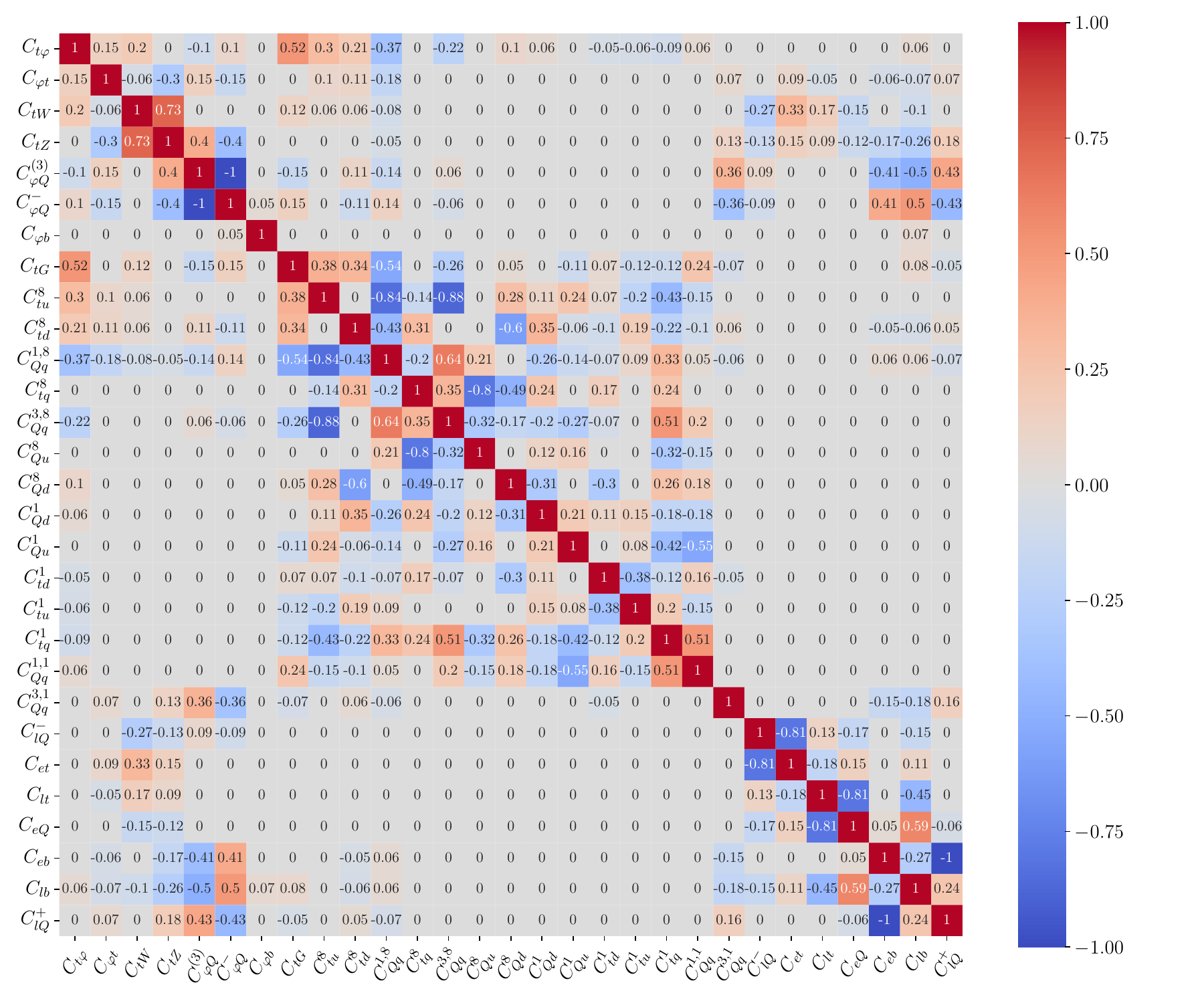}
\caption{\label{fig:corr_LHC_10}Correlation matrix obtained for the global fit including the data of the HL-LHC, Tevatron, LEP, the final stage of FCCee and a muon collider at 3 and 10 TeV.  Entries smaller than 5\% are set to zero. }
\end{figure}

\begin{figure}\centering
\includegraphics[width=1.1\columnwidth]{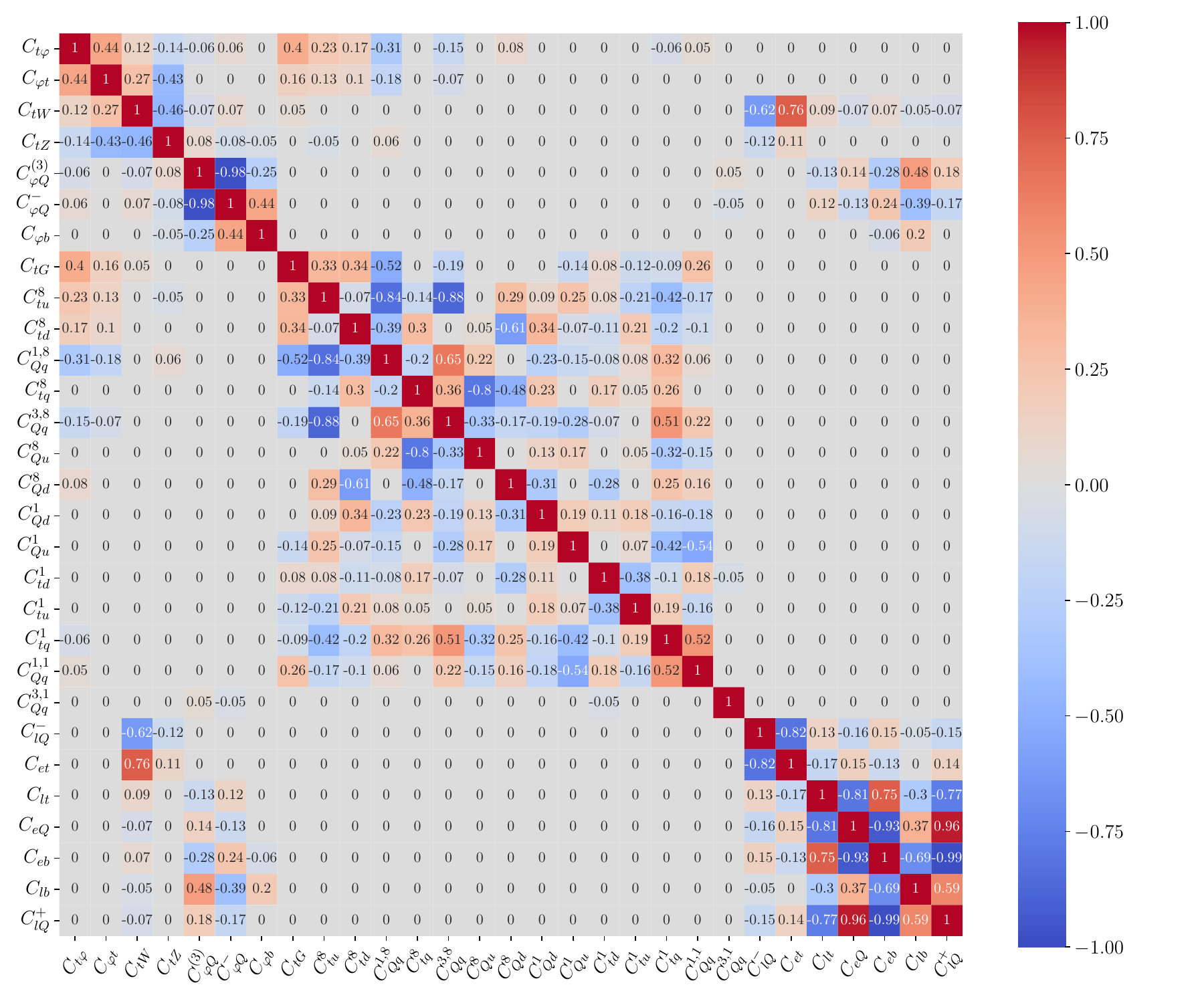}
\caption{\label{fig:corr_LHC_11}Correlation matrix obtained for the global fit including the data of the HL-LHC, Tevatron, LEP, the final stage of FCCee and a muon collider at 3, 10 and 30 TeV.  Entries smaller than 5\% are set to zero. }
\end{figure}